\documentclass[3p,number,preprint,times]{elsarticle} 
 
\usepackage{amssymb}
\usepackage{amsmath}
\usepackage{lipsum}
\usepackage[obeyspaces]{url}
\usepackage{hyperref}
\usepackage[utf8]{inputenc}

\usepackage{lineno}

\journal{Nuclear Instruments and Methods}

\begin{document}

\begin{frontmatter}



\title{Aerogel RICH Counter at the Belle II Detector}

\author[a,n]{I.~Adachi}
\author[f]{N.~Akopov}
\author[g]{D.~Augueste}
\author[g]{J.~Bonis}
\author[g]{L.~Burmistrov\fnref{fn1}}
\author[k]{S.~Dey}
\author[c,b]{R.~Dolenec}
\author[f]{G.~Ghevondyan}
\author[d]{R.~Giordano}
\author[b]{A.~Hvala}
\author[k]{T.~Iijima}
\author[h]{S.~Iwata}
\author[h]{H.~Kakuno}
\author[f]{G.~Karyan}
\author[i]{H.~Kawai}
\author[a]{T.~Kohriki}
\author[l]{T.~Konno}
\author[e,b]{S.~Korpar\corref{cor1}} \ead{samo.korpar@um.si}
\author[c,b]{P.~Kri\v zan}
\author[h]{S.~Kurokawa}
\author[a]{Y.~Lai}
\author[b]{A.~Lozar}
\author[b]{M.~Mrvar\fnref{fn2}}
\author[f]{G.~Nazaryan}
\author[a,m,n]{S.~Nishida\corref{cor1}} \ead{shohei.nishida@kek.jp}
\author[j]{S.~Ogawa}
\author[b]{R.~Pestotnik}
\author[b]{I.~Prudiiev}
\author[c,b]{L.~\v Santelj}
\author[b]{A.~Seljak}
\author[b]{L.~Senekovi\v c}
\author[a]{M.~Shoji}
\author[b]{K.~\v Spenko}
\author[h]{T.~Sumiyoshi}
\author[i]{M.~Tabata}
\author[a]{K.~Uno}
\author[a]{E.~Waheed\fnref{fn3}}
\author[h]{M.~Yonenaga}
\author[m]{Y.~Yusa}

\cortext[cor1]{Corresponding authors}
\fntext[fn1]{Present address: Universit\' e de Gen\` eve, Geneva, Switzerland.}
\fntext[fn2]{Present address: HEPhy, Vienna, Austria.}
\fntext[fn3]{Present address: Univ. of Melbourne, Melbourne, Australia.}

\affiliation[a]{organization={Institute of Particle and Nuclear Studies, KEK},
            city={Tsukuba},
            country={Japan}}
\affiliation[f]{organization={Alikhanyan National Science Laboratory},city={Yerevan 0036},country={Armenia}}
\affiliation[g]{organization={Universite Paris-Saclay, CNRS/IN2P3, IJCLab},city={91405 Orsay},country={France}}
\affiliation[b]{organization={Jožef Stefan Institute},
            city={Ljubljana},
            country={Slovenia}}
\affiliation[c]{organization={Faculty of Mathematics and Physics, University of Ljubljana},
            city={Ljubljana},
            country={Slovenia}}
\affiliation[d]{organization={University of Naples "Federico II" and INFN},
            city={Naples},
            country={Italy}}
\affiliation[e]{organization={Faculty of Chemistry and Chemical Technology, University of Maribor},
            city={Maribor},
            country={Slovenia}}
\affiliation[h]{organization={Tokyo Metropolitan University},
            city={Hachioji},
            country={Japan}}
\affiliation[i]{organization={Chiba University},
            city={Chiba},
            country={Japan}}
\affiliation[j]{organization={Toho University},
            city={Funabashi},
            country={Japan}}
\affiliation[k]{organization={Nagoya University},
            city={Nagoya},
            country={Japan}}
\affiliation[l]{organization={Kitasato University},
            city={Sagamihara},
            country={Japan}}
\affiliation[m]{organization={Niigata University},
            city={Niigata},
            country={Japan}}
\affiliation[n]{organization={The Graduate University for Advanced Studies (Sokendai)},
            city={Hayama},
            country={Japan}}

\begin{abstract}
We report on the design, operation, and performance of a novel proximity-focusing Ring Imaging Cherenkov (RICH) detector equipped with a multilayer focusing aerogel radiator, developed for the forward region of the Belle II spectrometer at the SuperKEKB $e^+e^-$  collider. The system achieves effective separation of charged pions, kaons, and protons across the full kinematic range of the experiment, from 0.5~GeV/$c$ to 4~GeV/$c$. To date, the detector has successfully operated in data taking, contributing to the collection and analysis of nearly 600~fb$^{-1}$ of Belle~II $e^+e^-$ collision data.

\end{abstract}




\end{frontmatter}


\section{Introduction}
\label{sec:introduction}

The Belle II experiment, located in Tsukuba, Japan, is dedicated to precision measurements of rare decays of $B$ and $D$ mesons and $\tau$ leptons. Following the highly successful operation of the original Belle spectrometer between 1999 and 2010, and numerous landmark physics results~\cite{physics-book}, the pursuit of possible deviations from the Standard Model in extremely rare decay channels requires a two-order-of-magnitude increase in the size of the data sample.

To achieve this goal, the KEKB $e^+ e^-$ collider underwent a major upgrade. The   SuperKEKB accelerator was constructed, and was designed to operate at event rates approximately 30 times higher than its predecessor and to ultimately deliver 50~ab$^{-1}$ of integrated luminosity~\cite{Ohnishi:2013fma}.  The substantially higher interaction rates and corresponding background levels required a comprehensive upgrade of the Belle spectrometer~\cite{Belle-II:2010dht,Adachi:2018qme,b2det-jinst}.

\begin{figure*}[hbt]
\begin{center}
\includegraphics[width=0.6\textwidth]{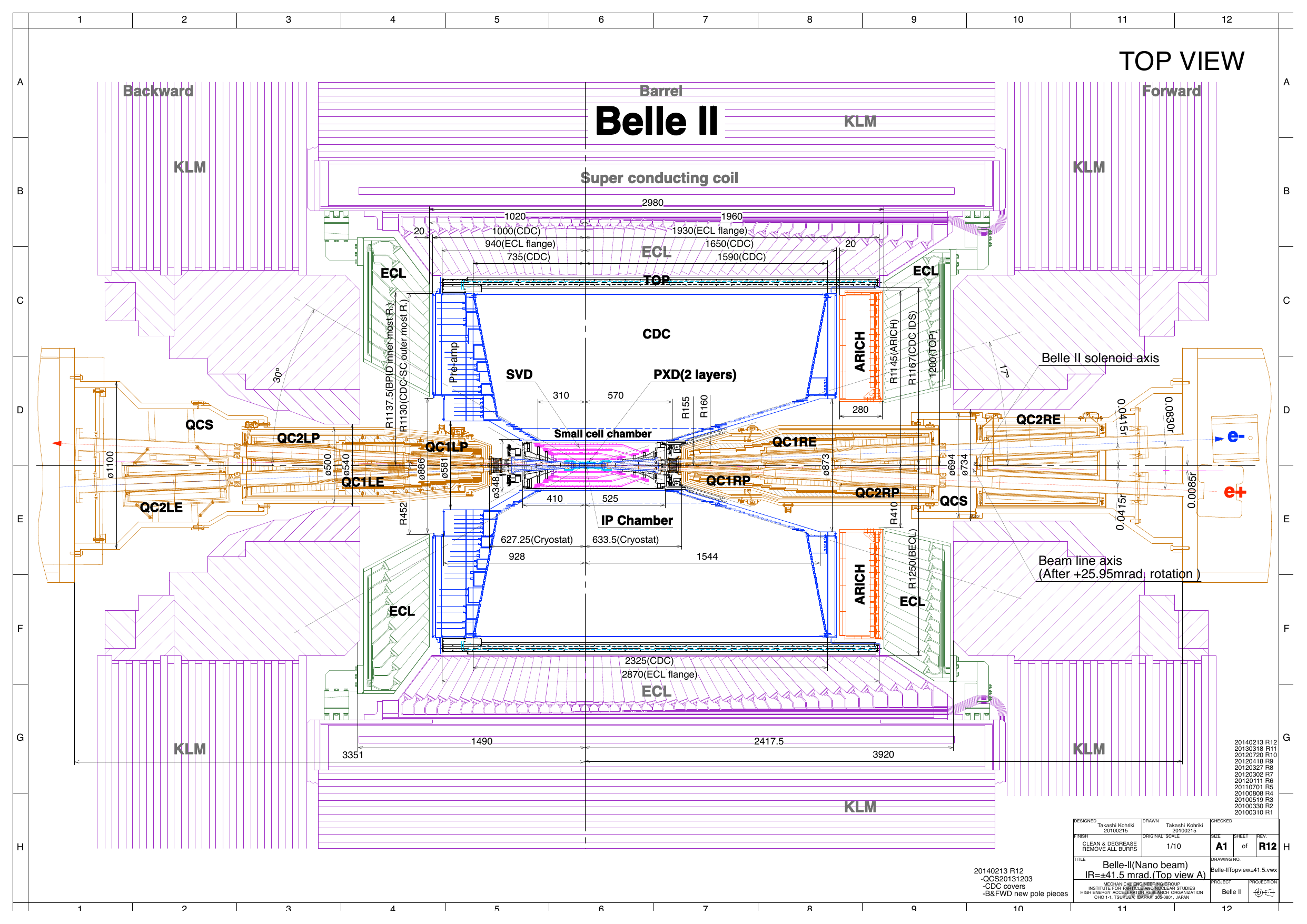}
\includegraphics[width=0.28\textwidth]{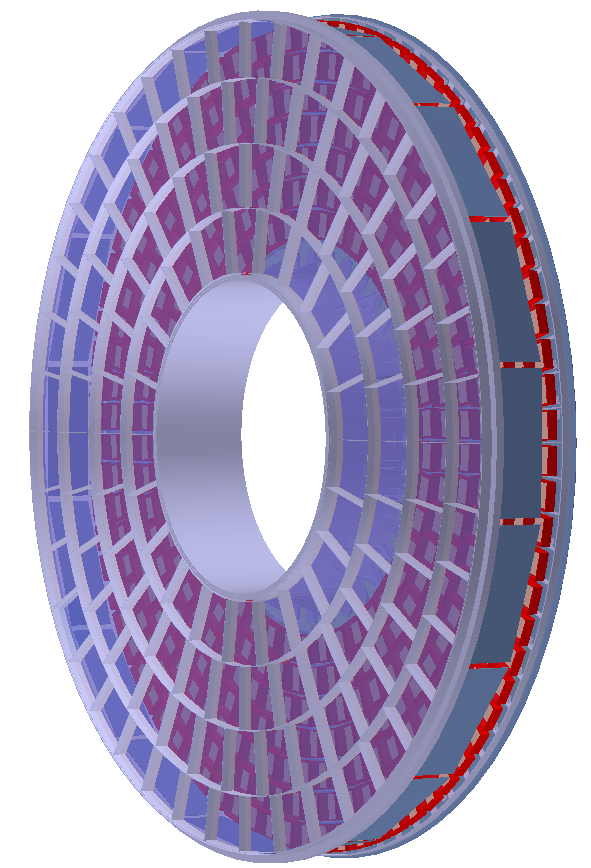}\end{center}
\caption{The Belle II spectrometer (left) and the geometry of the ARICH detector (right) with its main components, aerogel radiator (edges depicted in light gray) and photo-sensors (in dark red).
}
\label{fig:belle2-arich}
\end{figure*}

To meet its physics objectives, the Belle II experiment requires highly efficient kaon–pion separation for charged particles with momenta up to 4~GeV/$c$, achieved using two dedicated subsystems based on Cherenkov photon detection. Particle identification in the barrel region of the spectrometer is provided by the Time-of-Propagation (TOP) counter~\cite{Atmacan:2025jmh} while the Aerogel Ring Imaging Cherenkov (ARICH) detector—described in detail in this paper—was developed to distinguish kaons from pions across most of their momentum range in the forward (end-cap) region of the Belle II spectrometer (Fig.~\ref{fig:belle2-arich}).

The ARICH is a proximity-focusing Ring Imaging Cherenkov (RICH) counter (Fig.~\ref{fig:richprinc}) consisting of several key components:
\begin{figure}[hbt]
\begin{center}
\includegraphics[width=0.5\columnwidth]{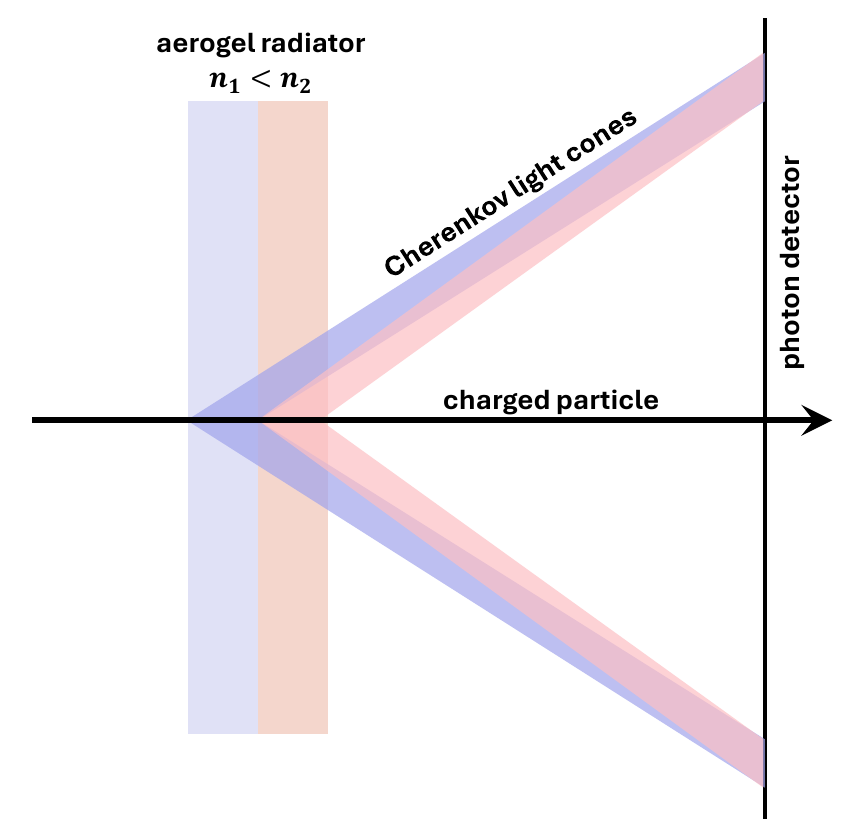}
\end{center}
\caption{Proximity focusing RICH with a dual radiator in the focusing configuration.
}
\label{fig:richprinc}
\end{figure}
a radiator, where charged particles emit Cherenkov photons; an expansion volume, allowing the photons to form ring images on the detection plane; an array of position-sensitive photon detectors, capable of single-photon detection in a strong magnetic field with high efficiency and excellent two-dimensional spatial resolution; and a readout system that records and processes the detected signals.

The remainder of this paper is organized as follows. Sections 2-9 discuss the design considerations and the main detector components, including the aerogel radiator, photon sensors, readout electronics, mechanical structure, and associated services. Section 10 outlines the simulation and reconstruction methods, and Section 11 presents the performance results obtained during operation with colliding electron and positron beams.

\section{Design choices}
\label{sec:design}

\begin{figure}[bt]
\begin{center}
\includegraphics[width=0.6\columnwidth]{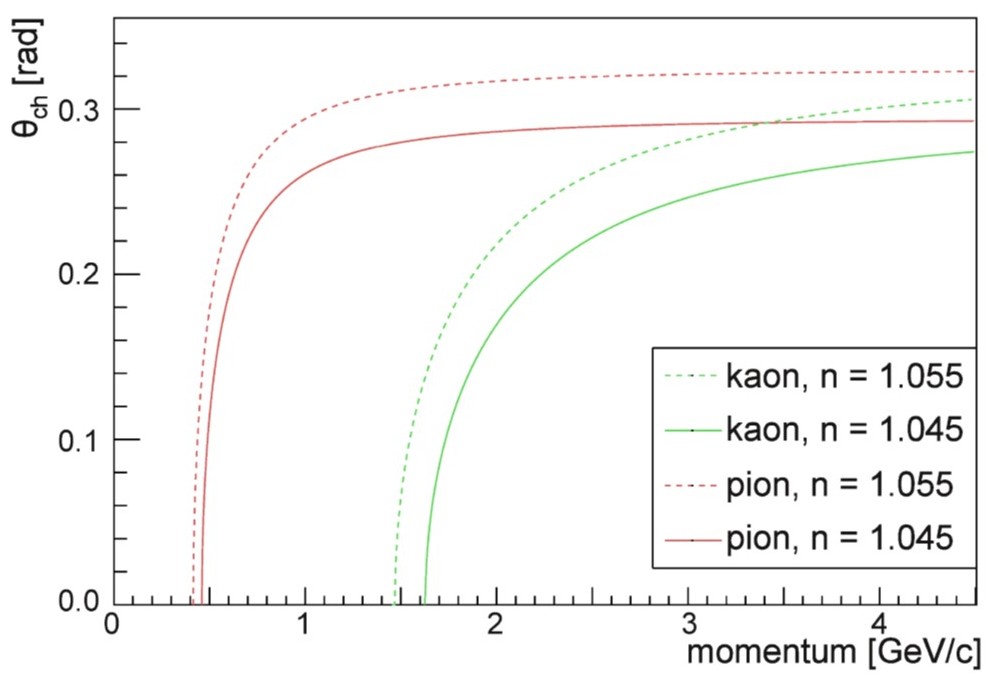}
\end{center}
\caption{Cherenkov angle for pions and kaons for two refractive indices of the Cherenkov radiator, 1.045 and 1.055. 
}
\label{fig:thetac-vd-p}
\end{figure}

The following considerations guided the design choices for the ARICH detector. The kinematic range of the Belle II experiment — covering pion and kaon momenta from 0.5~GeV/$c$ to 4~GeV/$c$ — requires the use of aerogel radiators with a refractive index of approximately 1.05 (Fig.~\ref{fig:thetac-vd-p}). 
To achieve the desired performance, a sufficient number of Cherenkov photons ($N \approx 10$)  must be detected for each ring image for at least one of the particle species. This requirement determines the total thickness of the aerogel radiator, which must be several centimeters.
  
The Cherenkov angle resolution required for $3 \sigma$ pion-kaon separation ($\sigma_{\rm track} \approx 7$~mrad)  can be achieved if the photon detector granularity is on the order of a few millimeters, given the space available for the expansion gap is about 20~cm.

A proof-of-principle prototype demonstrated excellent performance in both laboratory and beam tests~\cite{Matsumoto:2003wv}. However, two major challenges remained: increasing the number of detected Cherenkov photons and developing a single-photon detector capable of reliable operation in the 1.5~T magnetic field of the Belle II spectrometer. Both challenges were successfully addressed, as discussed below.
 
The key performance parameter of a RICH counter is the Cherenkov angle resolution per track, $\sigma_{\rm track} = \sigma_{\theta}/\sqrt{N}$, where $ \sigma_{\theta}$ is the single photon resolution and $N$ is the number of detected photons.
 Increasing the radiator thickness enhances photon statistics but degrades single-photon resolution in a proximity-focusing RICH due to emission-point uncertainty. Within the spatial constraints of the Belle II spectrometer, the optimal $\sigma_{\rm track}$ is achieved with a radiator thickness of about 20~mm~\cite{Matsumoto:2003wv,Iijima:2005qu}. However, in this configuration, the photon yield is insufficient for the required separation power.
 
This limitation can be overcome using a non-homogeneous, multilayer aerogel radiator~\cite{Iijima:2005qu,Krizan:2006pc,bib08:pk-hawaii,bib08:danilyuk}.  
By selecting different refractive indices for consecutive aerogel layers, the corresponding Cherenkov rings can be made to overlap on the photon-detector plane (Fig.~\ref{fig:richprinc})~\cite{Iijima:2005qu,Krizan:2006pc}. 

\begin{figure}[bt]
\begin{center}
\includegraphics[width=0.8\columnwidth]{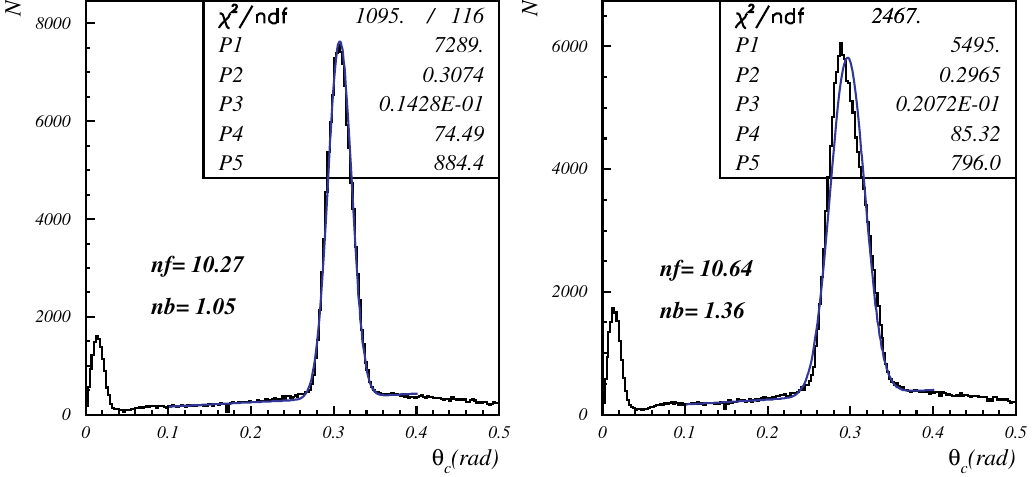}
\end{center}
  \caption{
 Proximity focusing RICH, proof of principle in a beam test:  the distribution of
    Cherenkov photon hits of the Cherenkov angle for a 4~cm homogeneous radiator (right) and for a focusing configuration with two 2~cm thick layers with $n_{1}$~=~1.046, $n_{2}$~=~1.056 (left)~\cite{Korpar:2007vxs}. 
  }
  \label{fig:richprinc-btest}
\end{figure}

This provides effective focusing of photons within the radiator, significantly reducing the spread arising from emission-point uncertainty. Such fine control of the refractive index is uniquely possible with aerogel, which can be produced with indices in the range $1.01-1.2$~\cite{bib08:aerogel-new}. 
Beam-test results comparing two 4 cm-thick radiators — one with uniform refractive index ($n = 1.046$) and the other with a dual-layer focusing configuration ($n_{1} = 1.046 , n_{2} = 1.056$)~\cite{Korpar:2007vxs} — clearly demonstrate the benefit of this approach. As can be seen in Fig.~\ref{fig:richprinc-btest}, the single-photon angular resolution improves from 
$\sigma_\theta=20.7$~mrad in the uniform case to 
$\sigma_\theta=14.3$~mrad for the dual-layer radiator, while maintaining a similar photon yield in both configurations~\footnote{An alternative geometry was also studied with an inverted order of the two aerogel tiles that would produce two separate rings~\cite{Iijima:2005qu}; while interesting, this de-focusing configuration was not further pursued because of ambiguities in the reconstruction and higher sensitivity to background.}.
The second primary design requirement is the selection of a photon sensor capable of detecting single photons in a strong magnetic field and withstanding the expected radiation load, corresponding to a 1 MeV neutron equivalent fluence of approximately $10^{12}$ n cm$^ {-2}$ and a gamma radiation dose of 100 Gy during the experiment's lifetime. This challenge was met with the Hybrid Avalanche Photodetector (HAPD) developed in collaboration with Hamamatsu Photonics, a proximity-focusing device selected as the baseline Cherenkov light sensor. Two alternative sensor technologies were also successfully evaluated: a multi-channel microchannel-plate photomultiplier tube (MCP-PMT)~\cite{mcp-bench,mcp-beam-test,mcp-magfield-timing,mcp-rich-tof} and a silicon photomultiplier (SiPM)-based photon detector~\cite{Korpar:2008zza,Korpar:2010zza}. 

\begin{figure}[tb]
\centering
    \includegraphics[width=0.5\columnwidth]{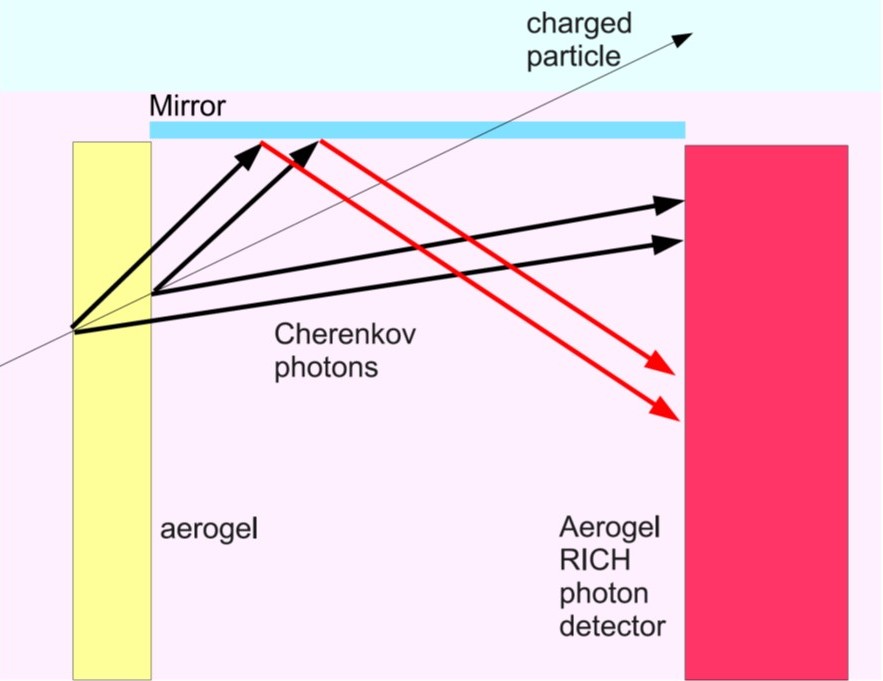}
\caption{Recovery of photons emitted at the outer edge of the ARICH detector acceptance: planar mirrors on the side wall of the ARICH vessel.
}
\label{fig:mirrors}
\end{figure}

Another important consideration in the detector design is the loss of photons at the side walls of the ARICH vessel (Fig.~\ref{fig:mirrors}). 
This effect is mitigated by installing planar mirrors that reflect photons back into the active region, thereby improving the overall Cherenkov light collection efficiency.

\section{Aerogel}

The radiator plane consists of two layers of wedge-shaped hydrophobic aerogel tiles with nominal refractive indices of $n = 1.045$ in the first layer and $n = 1.055$ in the second layer. As discussed in Sec.~\ref{sec:design}, they are chosen in such a way that Cherenkov rings from the first and second layers overlap on the detector plane (Fig.~\ref{fig:richprinc}) for the relevant kinematic range and a wide range of incidence angles~\cite{Krizan:2006pc}.

\begin{figure*}[tbp]
\begin{center}
\includegraphics[width=0.725\textwidth]{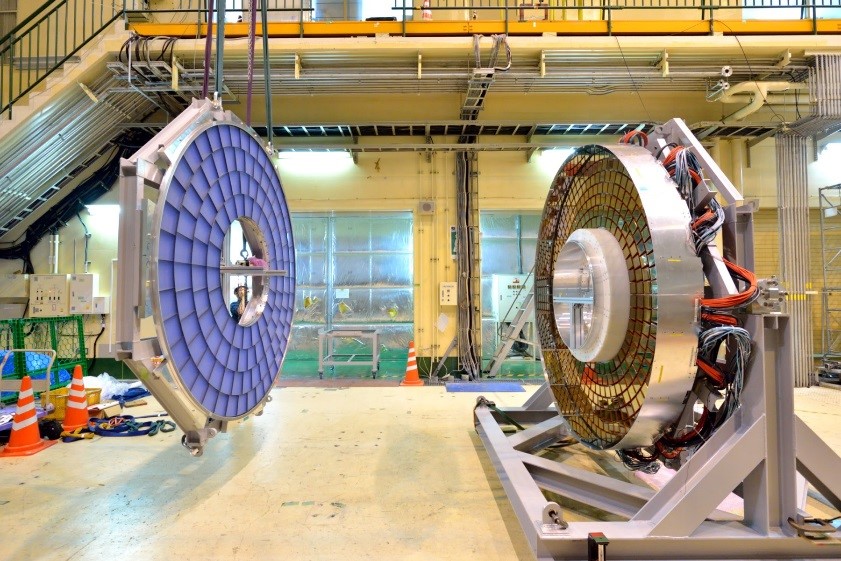}
\end{center}
\caption{Two halves of the ARICH detector: radiator plane covered with two layers of aerogel tiles (left), photon detector plane covered with 420 HAPDs and planar mirrors mounted on the side wall (right), before the two were joined to form a single vessel.
}
\label{fig:richlayout}
\end{figure*}

The aerogel tiles, totaling 124 for each refractive index,
 were fabricated using the super-critical drying method~\cite {Adachi:2017wye} at the Japan Fine Ceramics Center.
 As shown in Fig.~\ref{fig:richlayout}, the radiator system is segmented into four rings, each equipped with a separate type of wedge-shaped tiles.
The tiles were cut out of square-shaped $180 \times 180 \times 20$~mm$^3$
pieces of aerogel using a water-jet cutting device.

\begin{figure}[tbp]
\centering
    \includegraphics[width=0.4\columnwidth]{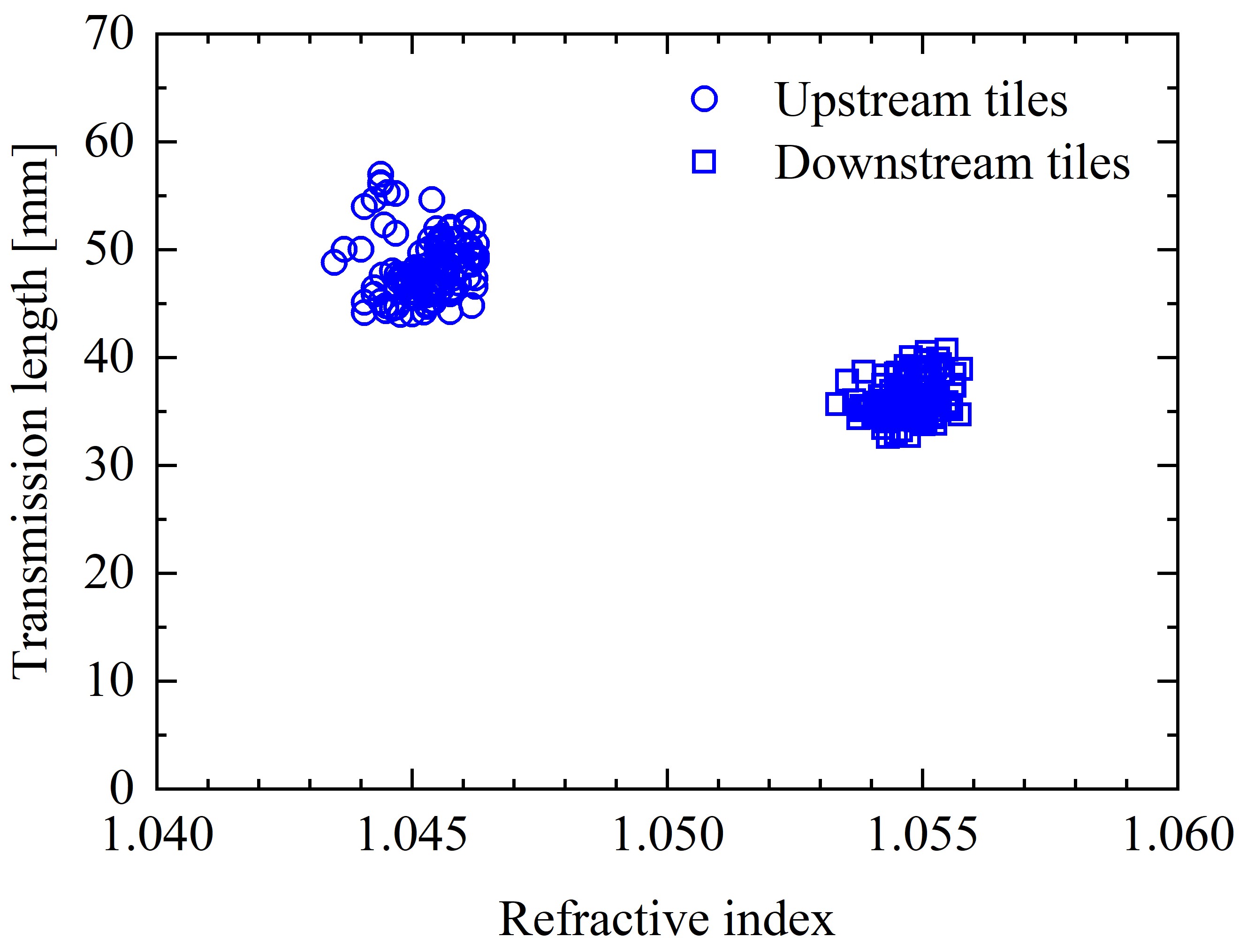}
    \includegraphics[width=0.45\columnwidth]{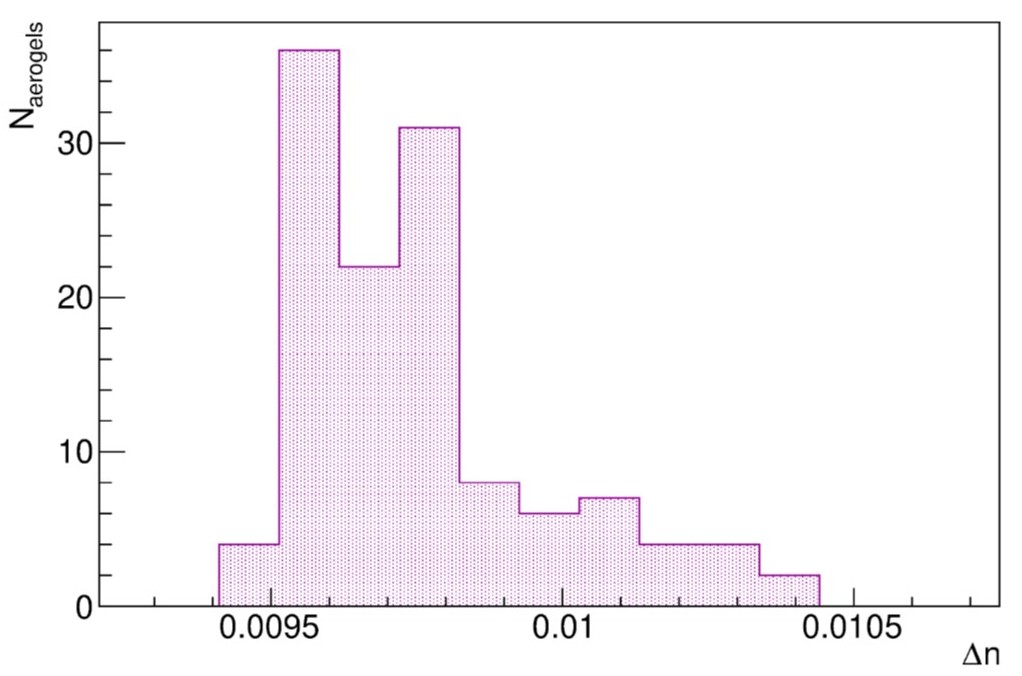}
\caption{Aerogel properties of the installed tiles. Left: 
the horizontal axis shows the refractive index, while
the vertical axis shows the transmission length at 400~nm. Right: the difference in refractive index between downstream and upstream tiles.}
\label{fig:aerogel}
\end{figure}

\begin{figure}[tbp]
\centering
    \includegraphics[width=0.49\columnwidth]{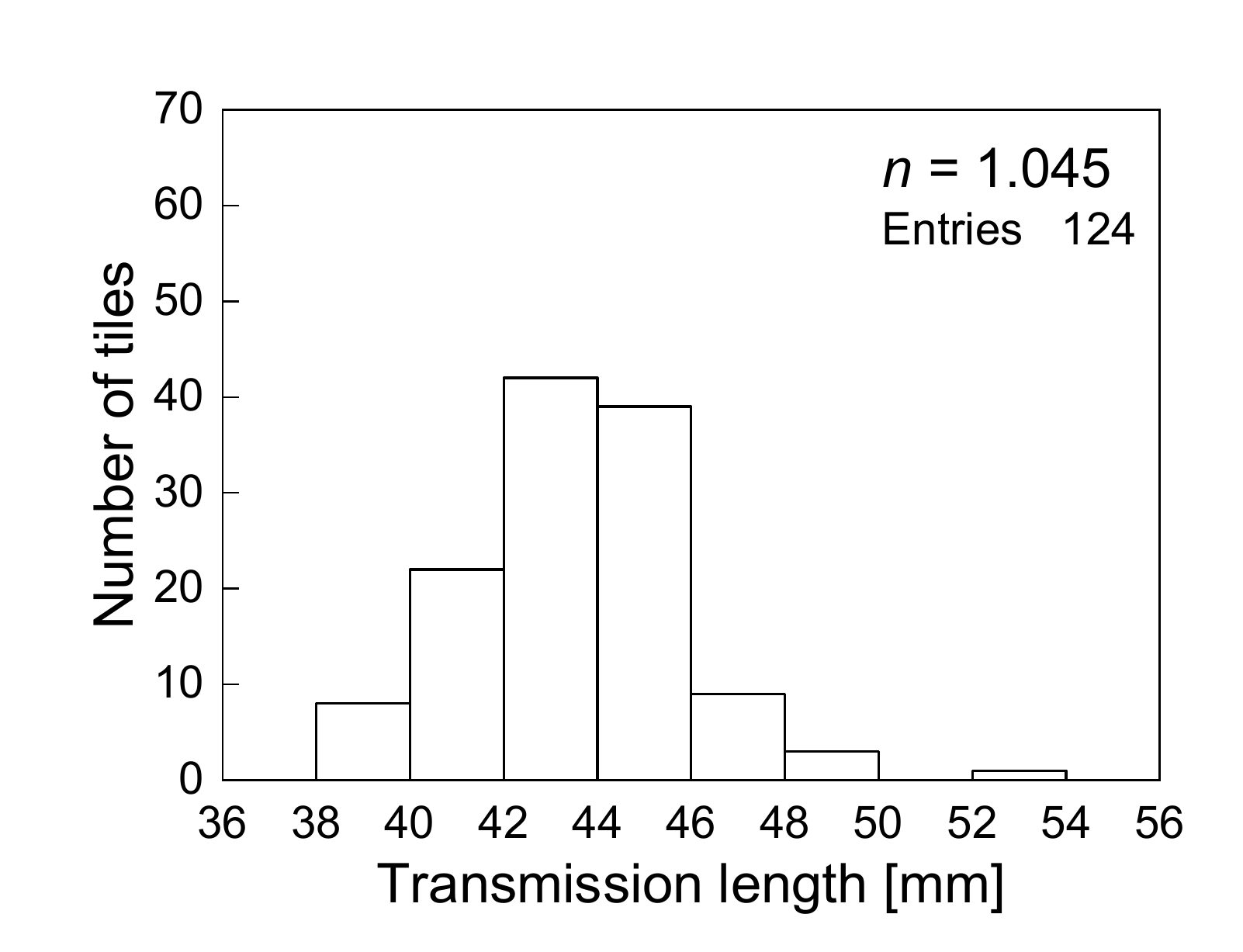}
    \includegraphics[width=0.49\columnwidth]{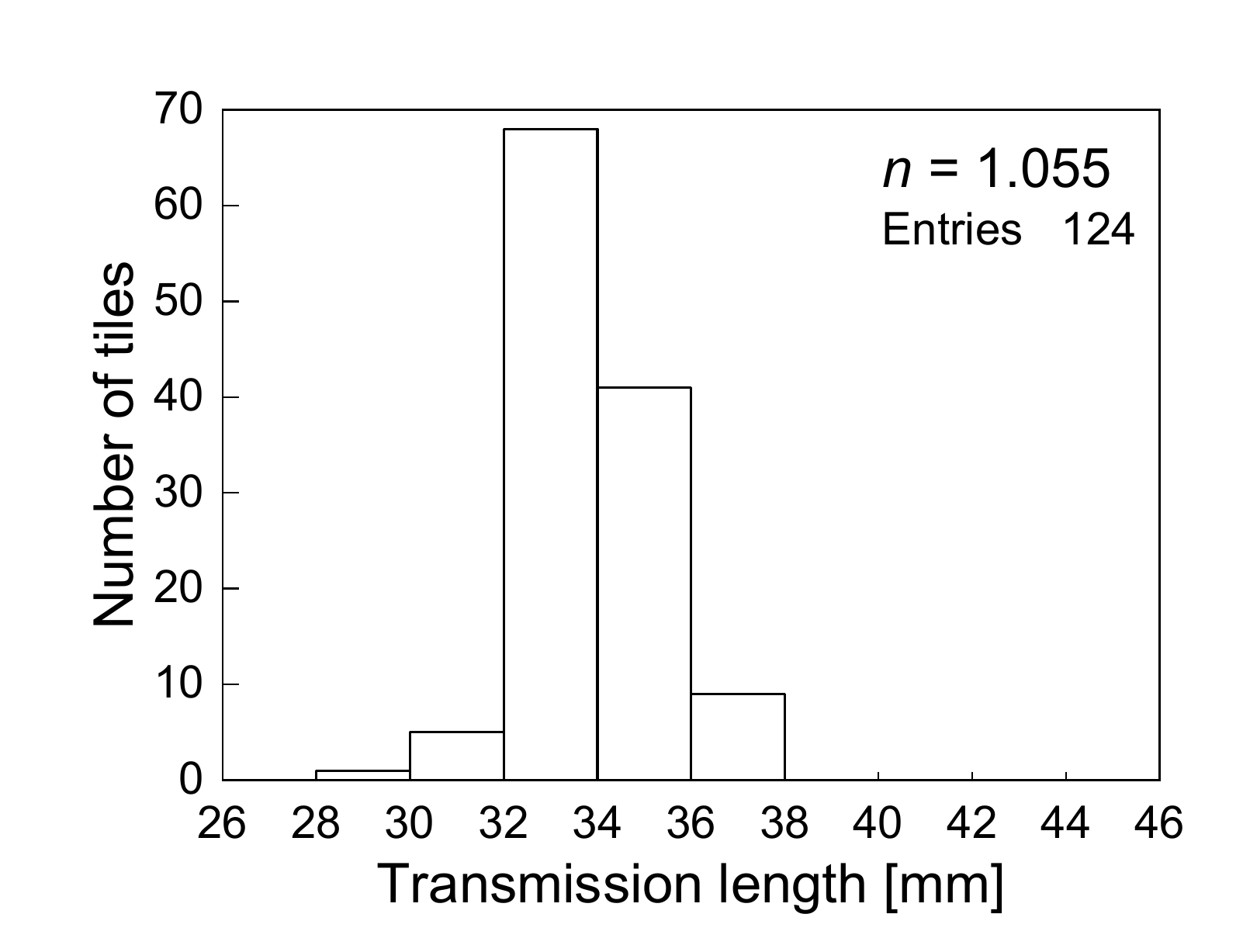}
\caption{Aerogel properties of the installed tiles: the distribution of transmission length at 400~nm for the upstream (left) and downstream tiles (right).
}
\label{fig:aerogel-transm-l}
\end{figure}

To achieve good PID performance, the aerogel tiles should be physically undamaged (with no cracks or chips)
and should have good transparency. 
Figures~\ref{fig:aerogel} and \ref{fig:aerogel-transm-l} show the distributions of the refractive indices and transmission lengths of the aerogel tiles from the mass production.
The refractive index of the tiles was controlled with appropriate precision so that the difference in the refractive index of the downstream and upstream tile pairs, $n_2-n_1=0.0098\pm0.0002$ (Fig.~\ref{fig:aerogel}), is well within the optimal interval~\cite{Krizan:2006pc}.


\section{Photon detector - HAPD}

\begin{figure}[tbp]
    \includegraphics[width=0.55\linewidth]{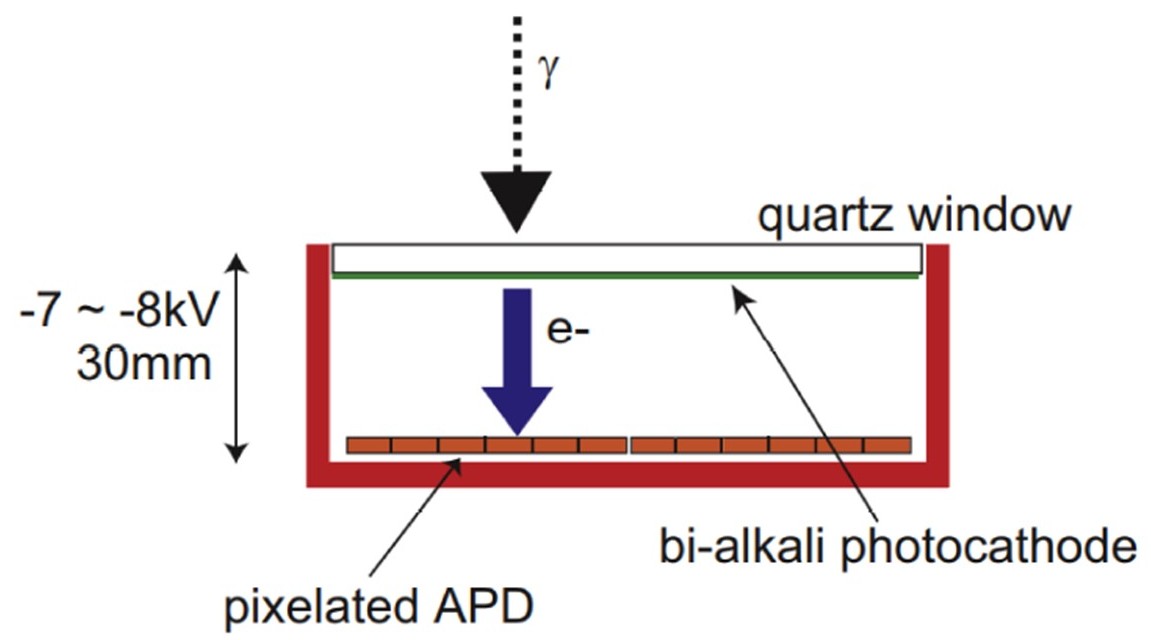}
    \centering    \includegraphics[width=0.35\linewidth]{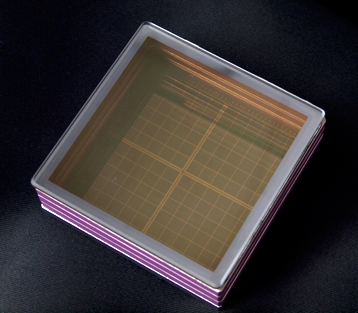}
\caption{HAPD single-photon sensor; left: the principle of operation~\cite{Nishida:2015qda,Korpar:2014bqa}, right: a photograph of the sensor.
}
\label{fig:hapd-principle}
\end{figure}

As the sensor for single Cherenkov photons, a hybrid avalanche photodetector (HAPD) (Fig.~\ref{fig:hapd-principle}) is used, where photoelectrons are accelerated in a static electric field and are detected with a segmented avalanche photodiode (APD).  The photon detector plane is covered with 420 HAPDs, arranged in seven concentric rings as shown in Fig.~\ref{fig:richlayout}.

\begin{figure}[tbp]
\centering
    \includegraphics[width=0.45\linewidth]{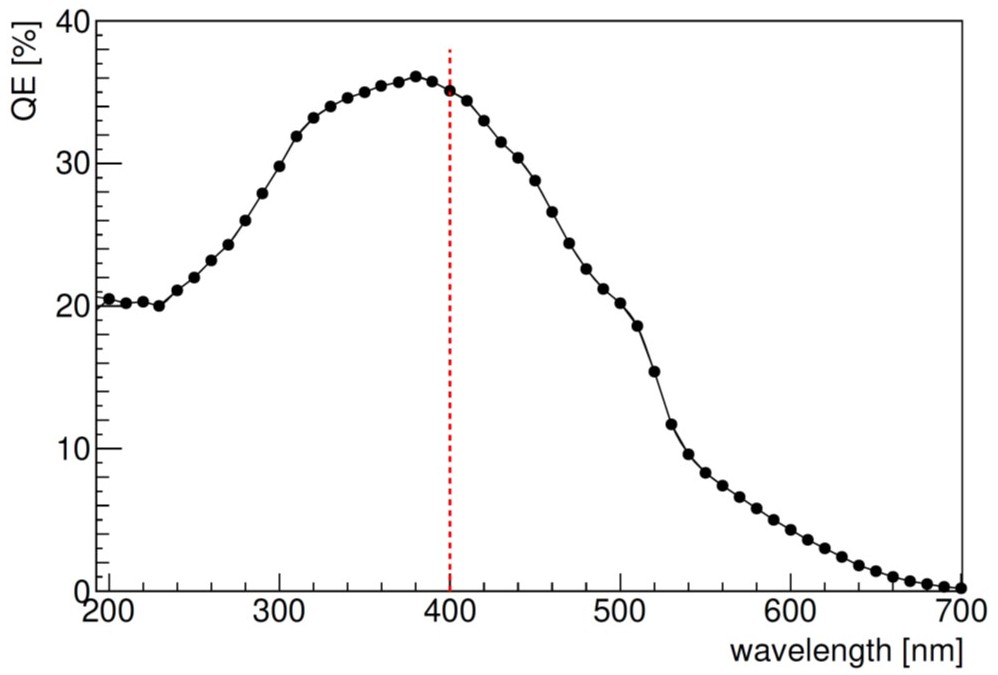}
    \includegraphics[width=0.45\linewidth]{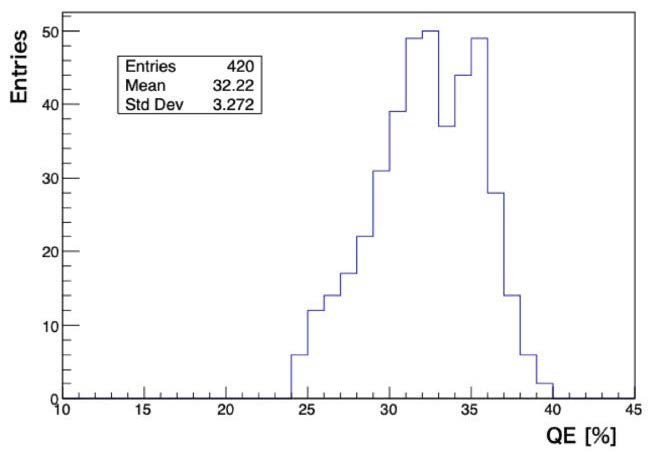}
\caption{HAPD single-photon sensor; left: quantum efficiency of a typical sensor as a function of wavelength, the vertical red line indicates the reference wavelength of 400~nm; right: quantum efficiency at 400~nm for all installed HAPDs.
}
\label{fig:hapd-qe}
\end{figure}
The HAPD employed in the Belle II ARICH detector was developed in collaboration with Hamamatsu Photonics~\cite{Nishida:2008zz,Korpar:2014bqa,Nishida:2015qda}. It has outer dimensions of $7.3 \times 7.3 \times 2.0$~cm$^3$, a bi-alkali photocathode on the inner side of the quartz window, and four pixelated avalanche photodiodes (APDs) at the bottom of the tube, each with 36 $5.1 \times 5.1$~mm$^2$ channels (Fig.~\ref{fig:hapd-principle}). The pad size is $4.9 \times 4.9$~mm$^2$, and the gap between pads is 0.2~mm within the APD; APDs are separated by 1.5~mm. The active area is about 65\% of the HAPD package size.
The total gain of approximately $3 \times 10^4$ results from the product of the bombardment gain and the gain of the APD. This gain is achieved through the application of a 6~kV high voltage to the tube together with an appropriate APD bias voltage, chosen to yield an APD gain of 30.
The quantum efficiency (Q.E.) of the photocathode at 400~nm is between $\approx25$\% and $\approx40$\% as can be seen in Fig.~\ref{fig:hapd-qe}. 

Following a set of radiation tolerance tests~\cite{Nishida:2014gra} with neutrons and gamma rays, sensor production was optimized (thicknesses of $p$ and $p^+$ layers, and an additional intermediate electrode). The final version of test samples retained the required performance at the neutron fluence and gamma radiation dose,  expected in the lifetime of the experiment. Another optimization of the sensor production (getter re-activation in the vacuum tube) was carried out to mitigate instabilities in the form of discharges when operated in the 1.5~T magnetic field~\cite{Santelj:2017epc}.  


The high-voltage system for HAPDs consists of 7 CAEN SY4527 crates, 45 CAEN A7042P 48-channel 500~V common floating return boards, which supply four bias voltages and one guard voltage for each of the 420 HAPDs, and 28 CAEN A1590 - AG590 16-channel 9~kV boards that supply 420 high voltages.

\section{Readout Electronics}

\label{sec:readout}
\begin{figure}[tbp]
\centering
   \includegraphics[width=0.4\linewidth]{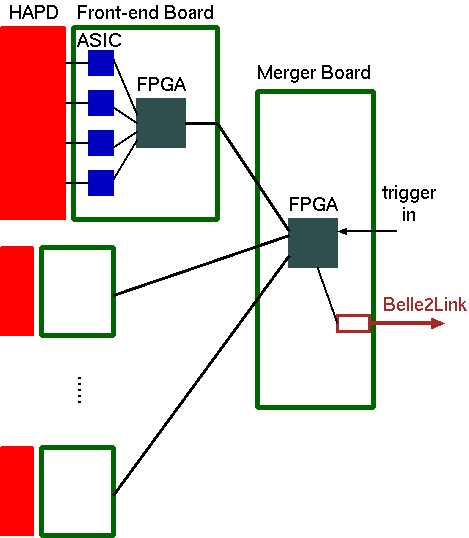}
\caption{Readout electronics for ARICH: schematics.}
\label{fig:hapd-readout}
\end{figure}

%
A total of 60480 readout channels are needed to equip 420 HAPDs,
 with a single bit of on/off hit information for each channel.
As shown in Fig.~\ref{fig:hapd-readout}, a front-end board (FEB)
with 4 ASICs and an FPGA (Xilinx Spartan 6, XC6SLX45-2FGG484)
is attached to each HAPD.
The 36-channel ASIC, named SA03, was developed in the X-FAB 0.35~$\mu$m process technology in collaboration with Japan Aerospace Exploration Agency (JAXA).

The ASIC  consists of a preamplifier with a variable gain between $17~\mathrm{mV/fC}$ and $56~\mathrm{mV/fC}$, a shaper with a shaping time that can be set to 170~ns, 210~ns, 240~ns, and 350~ns, and a
discriminator~\cite{Nishida:2012tua}.
It processes the analog signals from the HAPD. A common threshold voltage is applied to the discriminators of the ASIC,
while the baseline voltage of each channel can be adjusted
so that, effectively, the threshold of each channel can be set independently.

\begin{figure}[tbp]
    \centering
    \includegraphics[width=0.6\columnwidth]{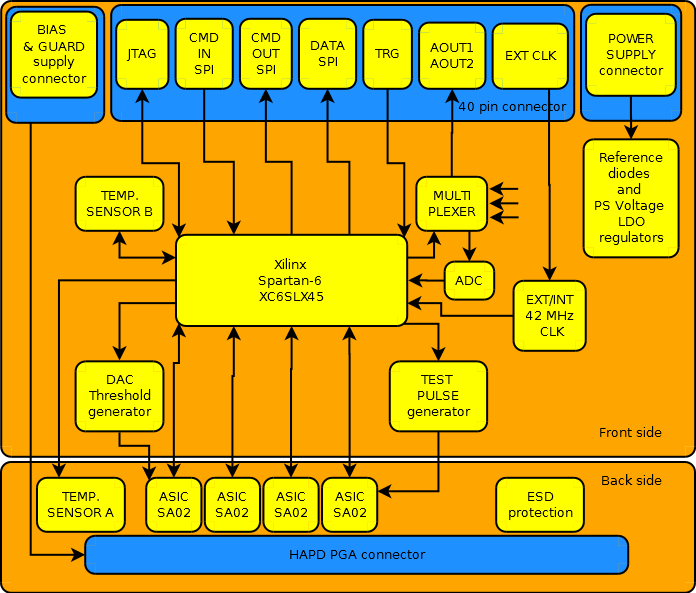}
    \caption{Functional schematic of the front-end board \cite{Pestotnik:2020feb} with four custom ASICs on the HAPD side, and a Xilinx Spartan-6 FPGA on the other side of the board.}
    \label{fig:feb}
\end{figure}

\begin{figure}[tbp]
\centering
    \includegraphics[width=0.6\linewidth]{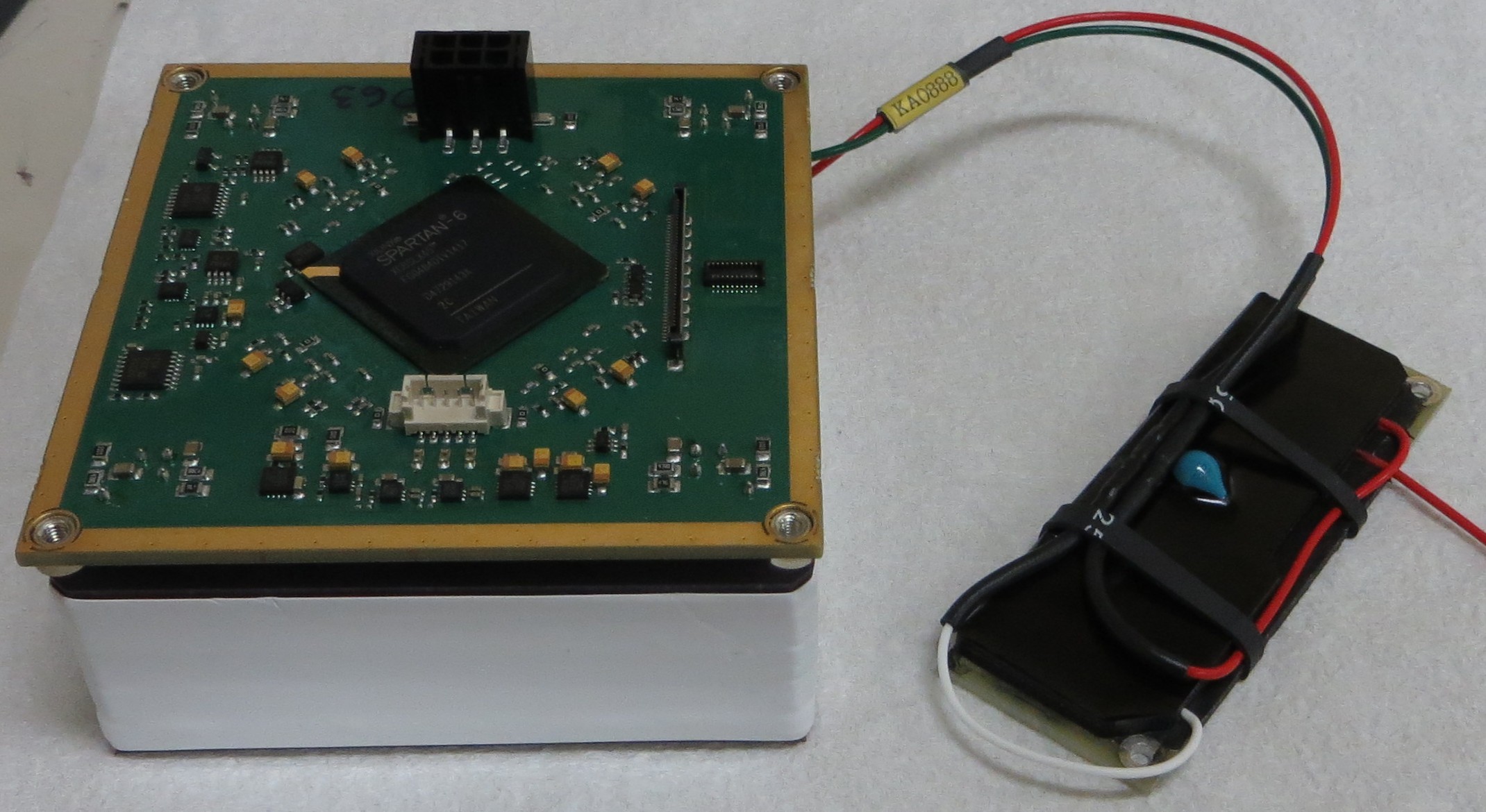}
\caption{HAPD readout and high-voltage supply: the front-end board mounted on top of the HAPD (with only the FPGA side visible) and the high-voltage supply board. When installed in the detector, the high-voltage supply board is mounted above the front-end board.}
\label{fig:hapd-hvboard}
\end{figure}

\begin{figure}[tbp]
\centering
   \includegraphics[width=0.44\linewidth]{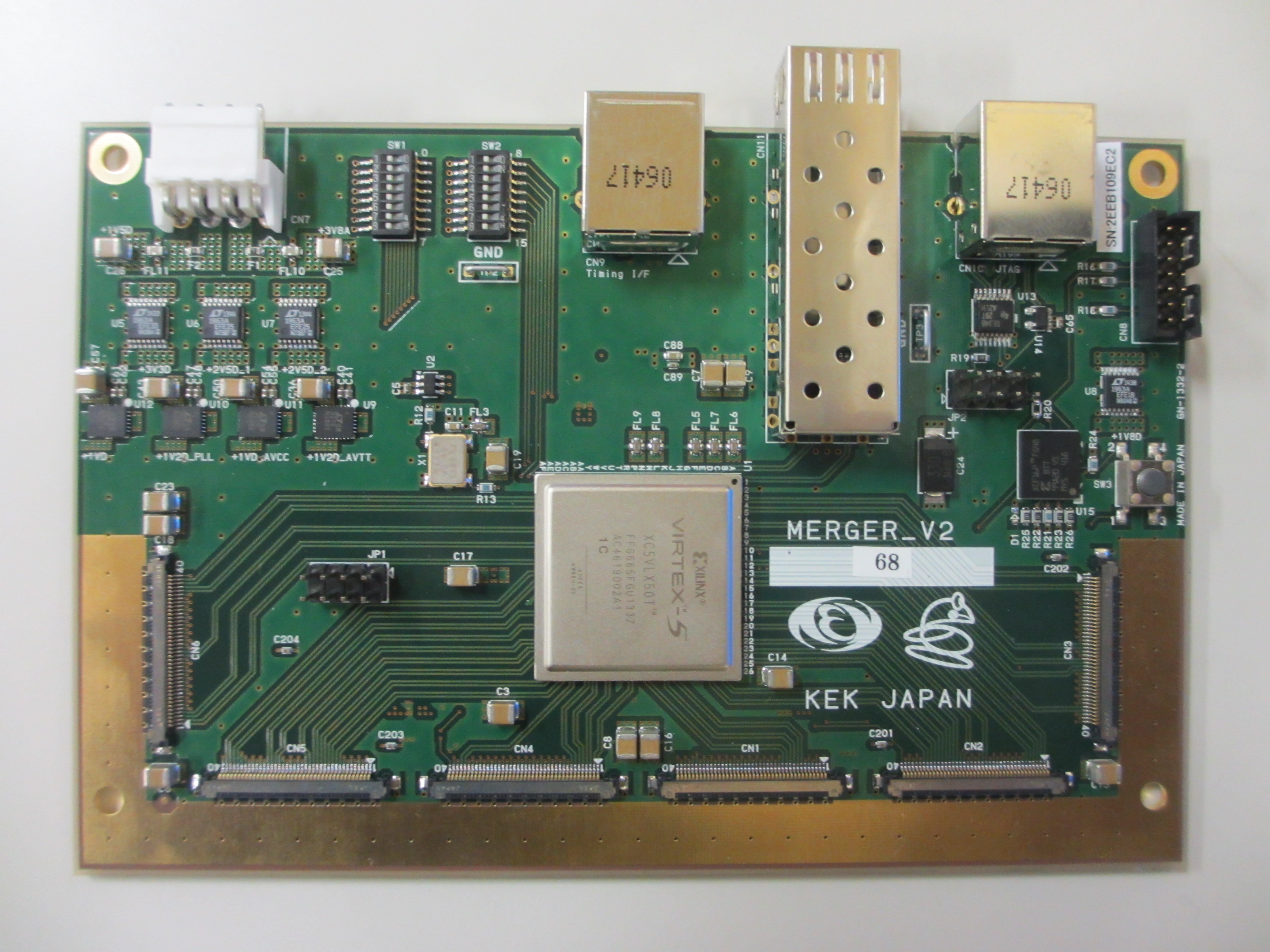}
   \includegraphics[width=0.5\linewidth]{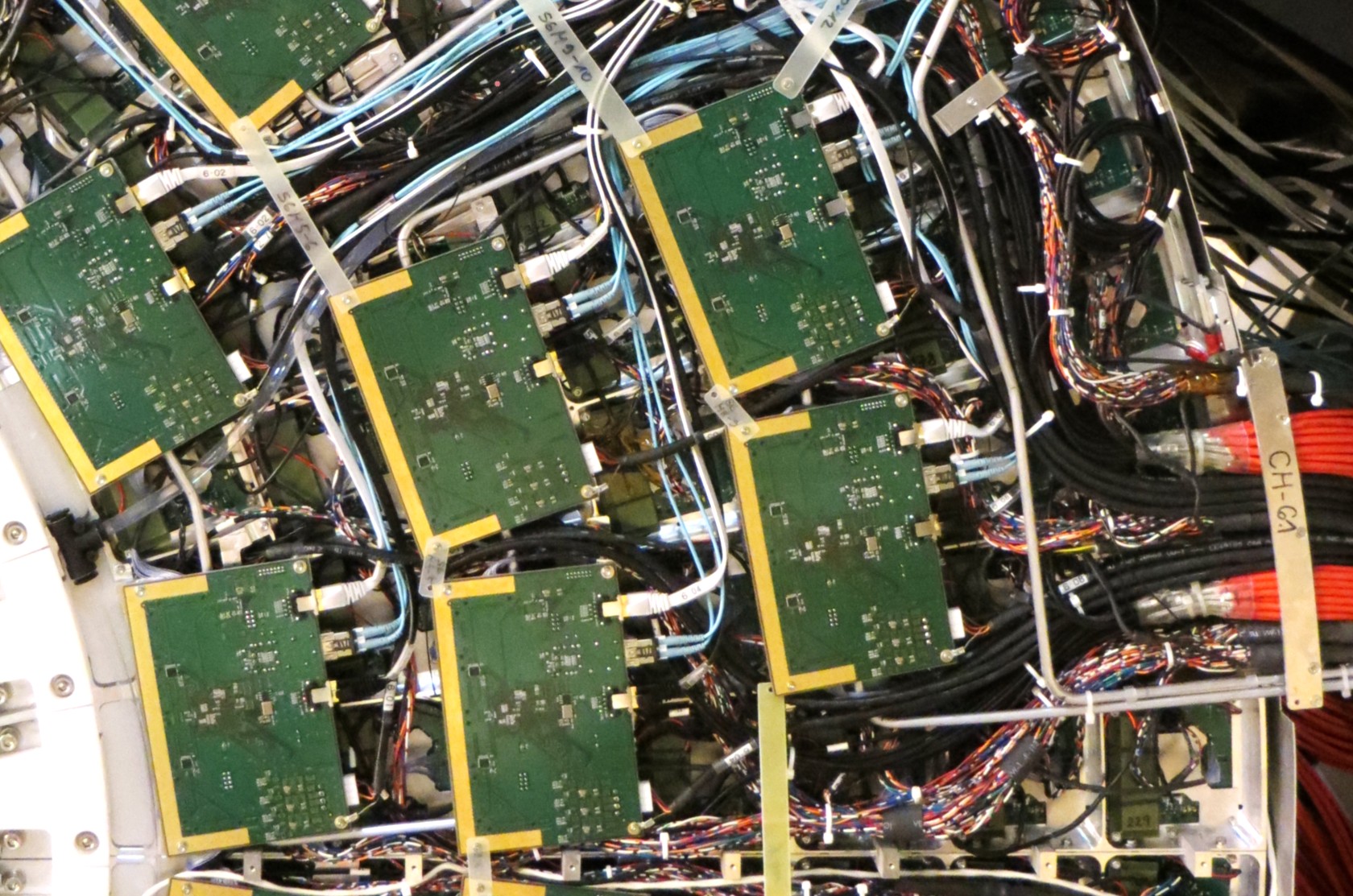}
\caption{Readout electronics: merger board (left) and mounted merger boards with services (right). 
}
\label{fig:hapd-backplane}
\end{figure}

The digital signals, output from ASICs, are processed in the FPGA~\cite{Nishida:2012tua,Seljak:2011zz}
(Figs.~\ref{fig:feb} and \ref{fig:hapd-hvboard}) and sent to the back-end electronics board ('merger') located behind the front-end boards
using parallel cables.
The merger board (Fig.~\ref{fig:hapd-backplane}), equipped with another FPGA (Xilinx Virtex 5, XC5VLX50T-1FFG665C), 
collects data from 5 to 6 front-end boards, performs zero suppression to reduce data size, and transfers the sparsified data to the central DAQ through the Belle2Link~\cite{b2link}.
The slow control, such as reading and writing of parameters in the ASICs, FEBs, and merger boards, is realized through
the Belle2Link. The merger also receives the first-level (L1) trigger signal and distributes it to the front-end boards.

The hit information from the HAPD sensors is read out in four adjacent time intervals. In the standard configuration, one timing bin corresponds to $125.6~\mathrm{ns}$ so that the time window for the readout is $502.4~\mathrm{ns}$. The middle two timing bins are adjusted to the correct timing with respect to the L1 trigger, and the two side bins are used to estimate the background level (see also Sec.~\ref{sec:operation}). 

The front-end readout board and merger board were designed to withstand a  1~MeV-neutron equivalent fluence of $10^{12}$ n/cm$^{2}$  and a gamma radiation dose of 100~Gy, as expected over the lifetime of the Belle II spectrometer. 

The low-voltage supply system for the read-out electronics consists of two Wiener MPOD systems and 12 low-voltage Wiener MPV8008LI modules (with 8 channels, floating output voltage 0~V-8~V at 5A, 40~W per channel, with less than 2~mVp-p ripple). 
These modules provide supply voltages of +3.8~V, +2~V, and $-2$~V to the 420 front-end boards and +1.5~V and +3.8~V to the 72 merger boards.
From the three supply voltages at a front-end board, a reference voltage of 1.25~V is generated by a diode, while other operating voltages are provided by low-dropout regulators (LDOs). For monitoring purposes, an analog-to-digital converter (ADC) digitizes the various internal voltage levels, which are selectable via a multiplexer. The  discriminator threshold level is adjusted using a digital potentiometer. The board is also equipped with an internal 42~MHz oscillator to clock the FPGA; an external clock is used during data-taking operation with the Belle II detector.

\subsection{Firmware Design}

The firmware system integrates a 64-bit instruction decoder, a data transmitter, a command receiver, and a response transmitter. Instructions are received via a unidirectional asynchronous Serial Peripheral Interface (SPI). The response to each command is returned encoded in a response word through a separate, dedicated one-way SPI channel. A trigger signal activates the data encoder, which then transmits data to the merger board via another one-way SPI.

Each hardware peripheral device is managed by a driver, which is activated by the instruction decoder. Additionally, a data register controls the mode of operation, sampling rates, and encoder frequencies. The unique FPGA device identifier (DNA) is used to identify the board.


To mitigate the impact of single-event upsets (SEUs) at the FPGAs of the FEBs, a scrubber of the configuration data of the FEBs is implemented in the firmware of the merger~\cite{Giordano:SEUmitiga}.
 The scrubber monitors the SEUs at FPGAs of FEBs and corrects them by a partial reconfiguration.

\section{Mechanical structure and services}

The ARICH detector components are mounted inside a donut-shaped vessel as shown in Fig.~\ref{fig:richlayout}. The vessel consists of two halves, one housing the aerogel radiator system, and the other the photon detectors with the read-out electronics. 
The total weight of approximately 400~kg is supported by two main mechanical components of the inner cylinder and the plate on which the HAPDs are mounted.

\begin{figure}[tbp]
\centering
    \includegraphics[width=0.7\linewidth]{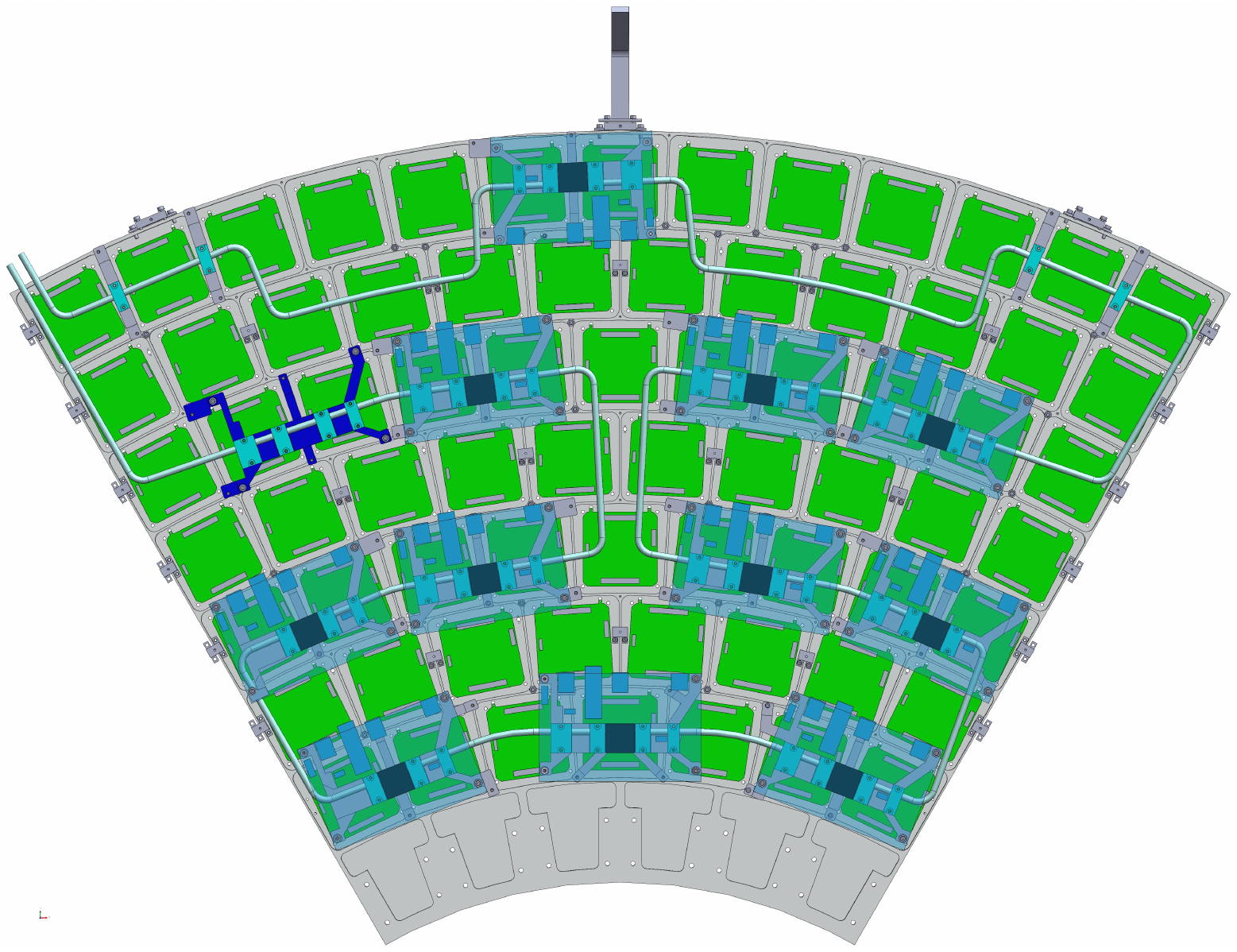}
\caption{Cooling system of the read-out electronics: cooling pipes are in thermal contact with the merger support elements that are in thermal contact with FPGAs (black squares) on the merger boards (transparent blue), merger board ground-plane, and the main aluminum support structure (in grey).  The cooling of the FE boards (in green) is provided through the thermal contact with the main aluminum structure. One of the merger boards was removed to reveal the associated merger support element (in dark blue). For other mergers, the PCB is displayed transparently, allowing the FPGAs to be visible in black and the connectors to be visible in blue.
}
\label{fig:cooling}
\end{figure}

\begin{figure}[tbp]
    \centering
    \includegraphics[width=0.7\columnwidth]{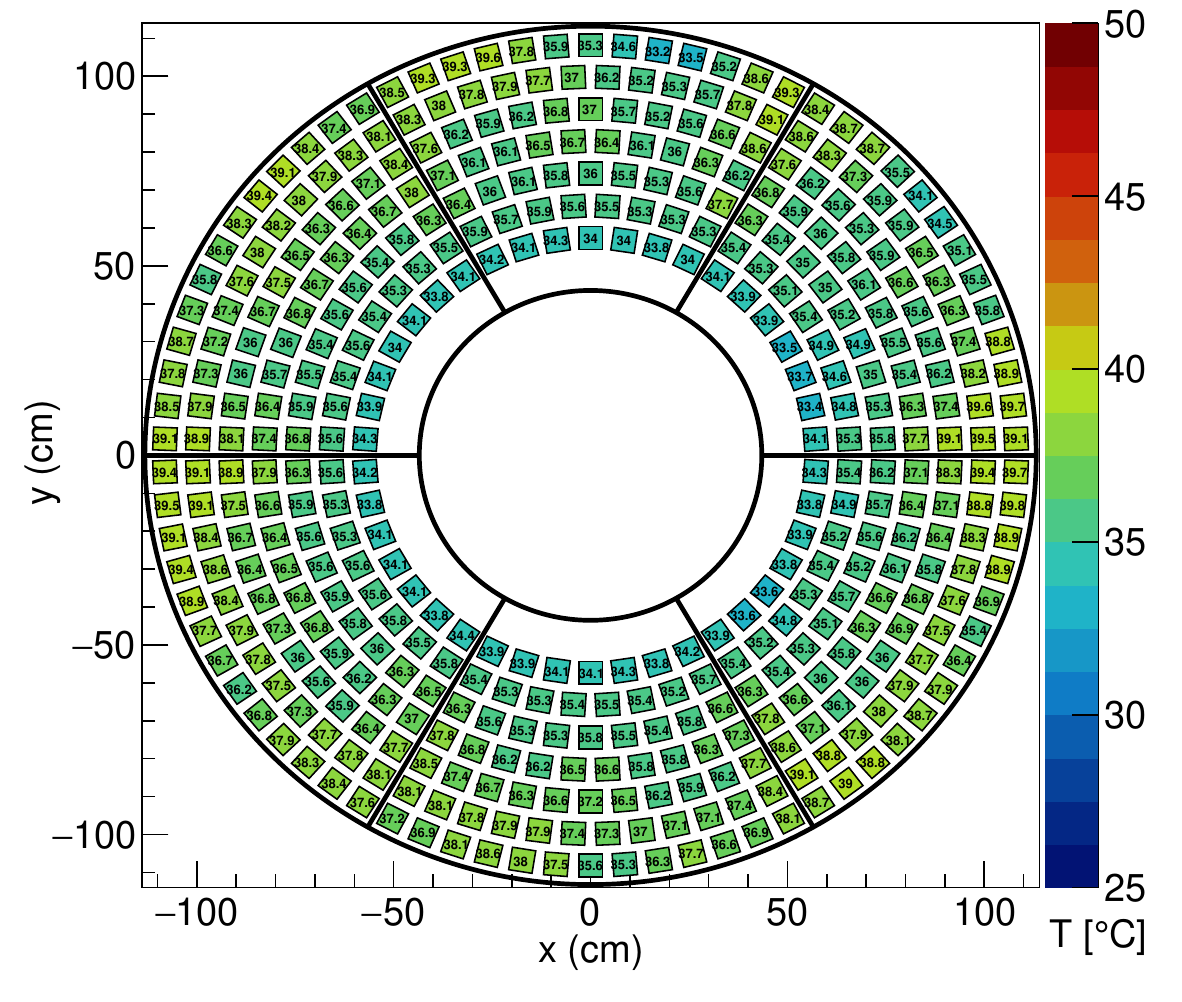}
    \caption{Temperature of the front-end boards during regular data taking operation.
    }
    \label{figure:temperature}
\end{figure}

On the photon detector side, the HAPDs and the associated read-out electronics are mounted on an aluminum support structure,  consisting of six azimuthal sectors. The support structure also provides a common electrical ground for the read-out electronics and sensors. It also conducts heat from the front-end boards to the cooling system.
Each of the six detector sectors is cooled by water flowing through a 5-m-long aluminum pipe with an outer diameter of 6~mm and 1~mm thick walls, as shown in Fig.~\ref{fig:cooling}. 
With a water flow of 1~l/min per sector and the output-input temperature difference of $\approx 2^{\circ}$C enough cooling is provided to extract $\approx 130$~W of heat from each of the sectors. With this arrangement,  the temperature of the front-end boards is maintained at approximately  40$^{\circ}$C (Fig.~\ref{figure:temperature}).


On the inside of the outer wall of the ARICH vessel, 18~planar mirrors are mounted to reflect the photons that would otherwise be lost back to the photo-sensitive area (Fig.~\ref{fig:mirrors}), as discussed in Sec.~\ref{sec:design}. The mirrors are 13.1~cm wide and 37.3~cm long, with a reflectivity exceeding 85\%  in
the wavelength range between 250~nm and 600~nm. The front surface technology is
used to reflect photons from the surface coating, to prevent the emission of additional
Cherenkov photons in the glass.

\section{Construction}

\subsection{Quality assurance of the components}

All ARICH components underwent several quality assessment (QA) measurements before assembly. The operational parameters of the photo-sensors were determined, including the operating values of high voltage and APD bias voltages, as well as measurements of leakage currents, quantum efficiency, and a performance test in a magnetic field. Voltages and currents on the front-end electronics were monitored, and the response of each channel was tested. Currents through high voltage distribution boards were measured with an applied voltage of 9~kV. If all components had a satisfactory quality and response, the final module, consisting of an HAPD, a front-end board, and a high-voltage distribution board (Fig.~\ref{fig:hapd-hvboard}), was assembled and re-tested.

\begin{figure}[tbp]
    \centering
    \includegraphics[width=0.7\columnwidth]{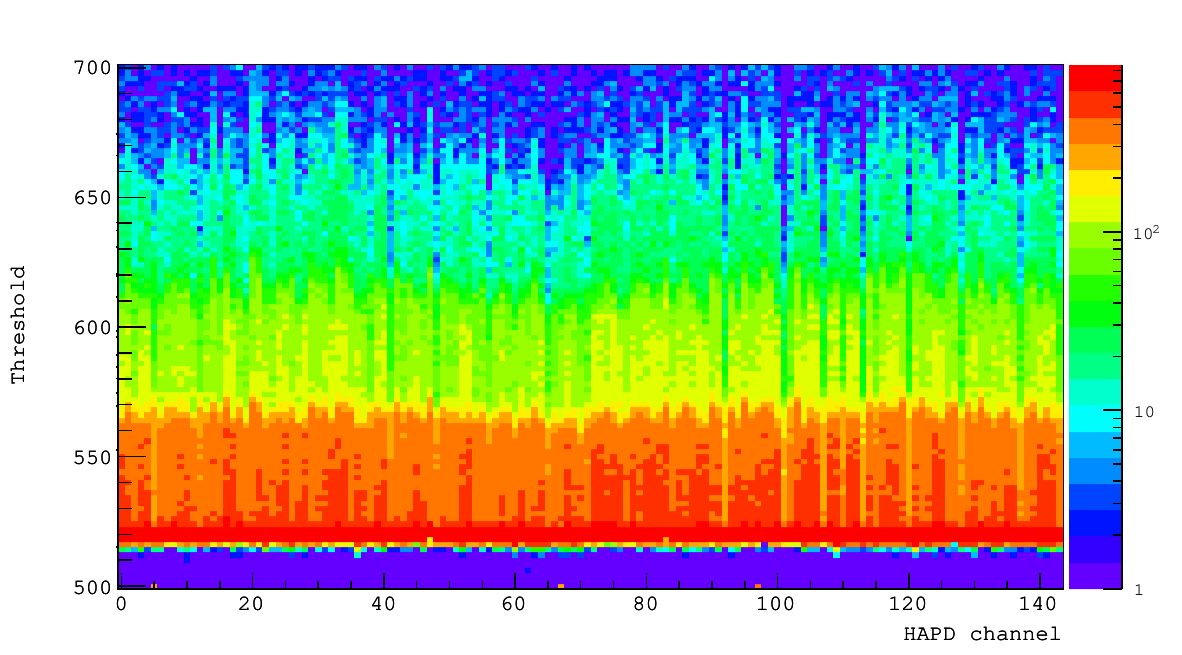}
    \caption{Threshold scan of the 144 channels of a HAPD: the responses of pads illuminated by short laser light pulses at different threshold voltages. }
    \label{figure:thscan:laser}
\end{figure}

We scanned the sensor surface over the channel centers and measured its response to short, low-intensity laser pulses at various discriminator values (Fig.~\ref{figure:thscan:laser}).
From this measurement, the gain of individual channels was extracted. 

We also measured and monitored the refractive indices of aerogel tiles and their transmission lengths (Figs.~\ref{fig:aerogel-transm-l} and \ref{fig:aerogel}). 

\subsection{Construction}

During the quality assurance of the photosensors, we observed that Q.E. varies significantly from sample to sample, as shown in Fig.~\ref{fig:hapd-qe}; the variation is significant, as the production specification required only the minimal quantum efficiency value. 
\begin{figure}[tbp]
\centering
     \includegraphics[width=0.65\linewidth]{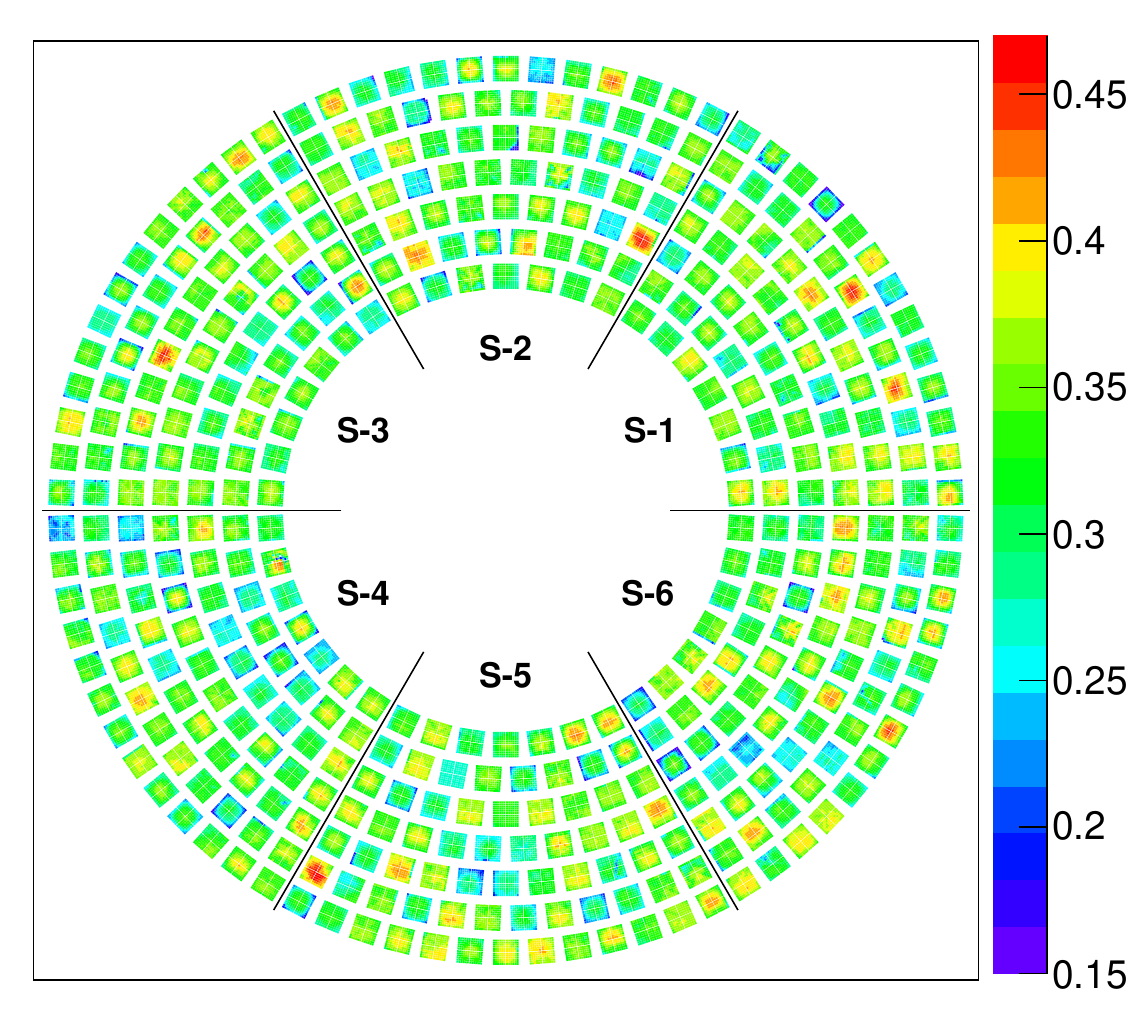}
\caption{ARICH photon detector plane: quantum efficiency at 400~nm as a function of the photon impact coordinate. 
}
\label{fig:hapd-qe-fulldet}
\end{figure}
To equalize the ARICH detector performance across the whole active surface, we randomly selected the mounting positions of the HAPDs as displayed in Fig.~\ref{fig:hapd-qe-fulldet}. 
\begin{figure}[tbp]
\begin{center}
\includegraphics[width=0.4\columnwidth]{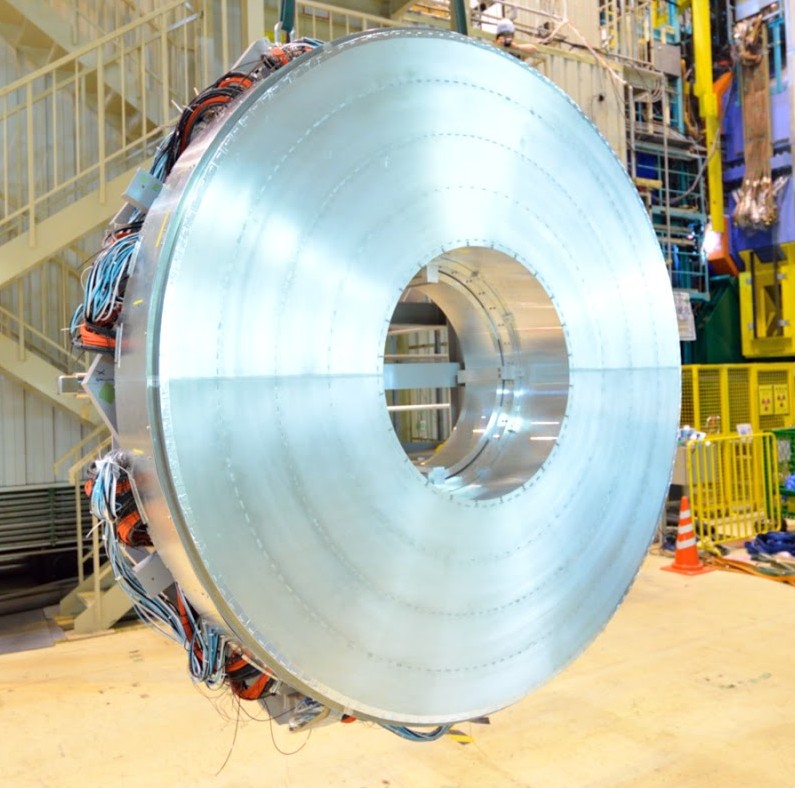}
\end{center}
\caption{ARICH detector construction: combining the two detector halves. 
}
\label{fig:arich-constr}
\end{figure}
%


The aerogel tiles were mounted on one of the two halves of the ARICH mechanical structure. They were positioned at the centers of tile slots formed by 1~mm-thick aluminum strips, with an additional 1~mm gap between each tile edge and the aluminum on all four sides, resulting in a total separation of 3~mm between adjacent aerogel tiles. The tile slots were lined with black paper on the upstream side and along the side walls. To ensure stability when the detector is vertically oriented, the tiles were secured to the ARICH mechanical structure using very thin carbon-fiber strings.
The photodetector modules were mounted to the aluminum structure of the other detector half. The cables for power supply and data transfer were custom-made primarily on-site. The eighteen planar mirrors were installed on the sides of the detector plane. A borated polyethylene neutron shield was mounted on the inside of the ARICH vessel, surrounding the inner wall (in white in Fig.~\ref{fig:richlayout}). Finally, the two halves of the ARICH detector were combined, and aluminum panels were attached at the edges, as seen in Fig.~\ref{fig:arich-constr}.

\section{Slow control system}
The Belle II ARICH slow control system~\cite{Pestotnik:2023slc} consists of four subsystems. The High Voltage System controls and monitors the HAPD high voltages, the Low Voltage Control System manages the voltage supplies to the readout electronics, the Environmental Monitor monitors the detector temperature, and the Front-End Board Control System uploads firmware, sets parameters of the readout chip, controls temperature, and manages the single event upset mitigation controller~\cite{Giordano:SEUmitiga}.

\begin{figure}[tbp]
    \centering
    \includegraphics[width=0.7\columnwidth]{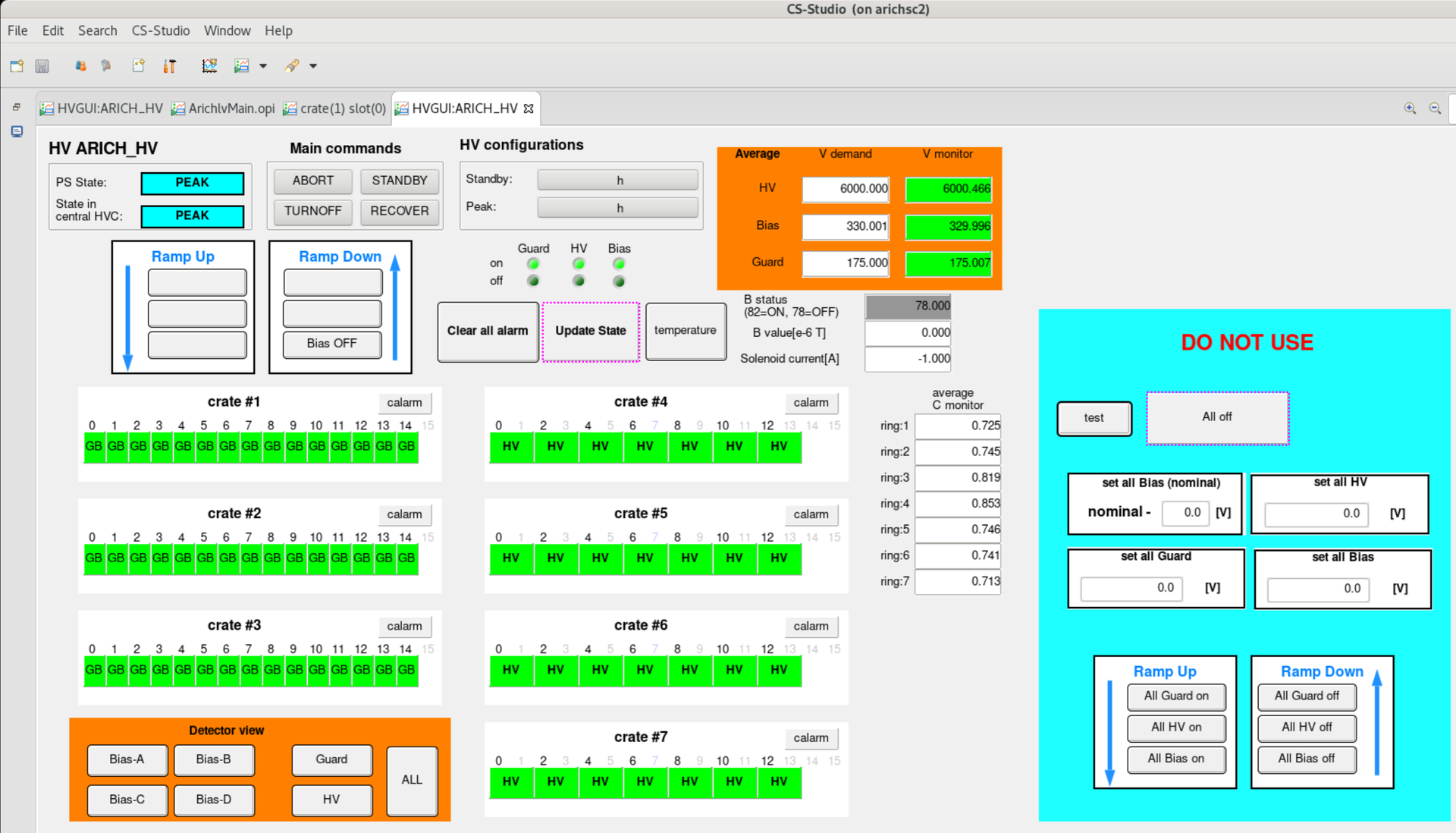}
    \caption{The main graphical user interface for slow control}
    \label{figure:slc-gui}
\end{figure}

The control daemons communicate with other processes using the common Belle II Belle2Link~\cite{b2link} and the Network Shared Memory 2 (NSM2) protocol. They accept requests to enable and disable supply channels and adjust hardware settings. Configuration settings are loaded from a common Belle II database, allowing for flexible and controlled value changes. The slow control system also continuously monitors voltage, current, and other detector parameters, such as temperatures and the number of hits. The values of the monitored parameters are regularly stored in the EPICS Archiver Appliance of the Belle II experiment \cite{EPICSArchiver}, and only significant changes from previous readings are saved. This approach reduces the need to store large amounts of data from over 15,000 monitoring variables. The slow control graphical user interfaces implemented in Control System Studio \cite{Clausen2007CSS,arich:css} visualize the current status and history of the slow control variables in an organized manner (Fig.~\ref{figure:slc-gui}). 
Finally, we note that a special tool was developed with a graphical interface that visualizes the connections between different detector parts. This feature was crucial during the installation and commissioning phases as well as during operation for identifying malfunctioning parts. 

%

\subsection{High Voltage and Low Voltage Systems}

\begin{figure}[tbp]
    \centering
    \includegraphics[width=0.75\columnwidth]{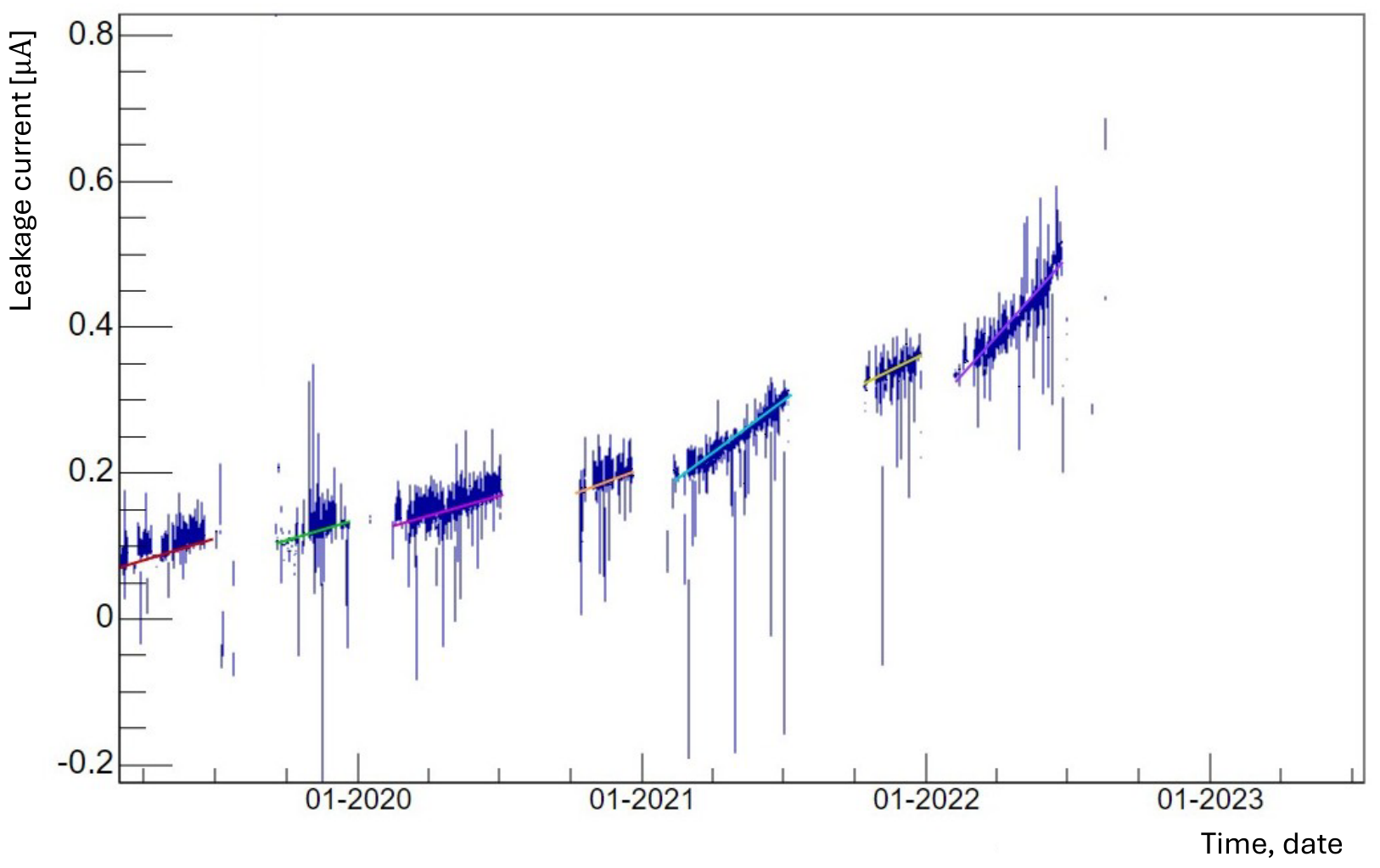}
    \caption{Dark current of one of the 36-channel APDs (one quarter of the HAPD), shown as a function of time; adopted from \cite{dark-current}.
    }
    \label{figure:bias}
\end{figure}

The high-voltage (HV) system is controlled by HV daemons that communicate with the hardware using the CAEN HV wrapper library. To minimize discharge risks, all 6~HV channels supplying a given HAPD follow well-defined transitions between different system states. Hardware interlocks ensure safe operation. Over 10,000 parameters are read from the high-voltage boards every ten seconds and recorded in the archiver database. For instance, the history of bias currents helps to estimate background irradiation levels on various parts of the detector (Fig.~\ref{figure:bias}).

The low-voltage supply slow control system is controlled by a separate deamon that communicates with the Wiener MPOD crate controller using the Simple Network Management Protocol (SNMP). In the same manner as the high voltage deamon, it ensures that the groups of voltages supplying a group of HAPD baseboards and merger boards are switched together.


\subsection{Data quality monitors}

Additional processes continuously extract key parameters from the reconstruction running online on a fraction of the data. These parameters include the Cherenkov angle of high momentum tracks, the number of hits per track, the number of hot and dead channels, and the temporal distribution of hits. The temporal changes of these parameters are monitored via various web interfaces (Figs.~\ref{figure:masked} and \ref{figure:perf-vs-time}).

\subsection{Environmental monitors}

\begin{figure}[tbp]
    \centering
    \includegraphics[width=0.5\columnwidth]{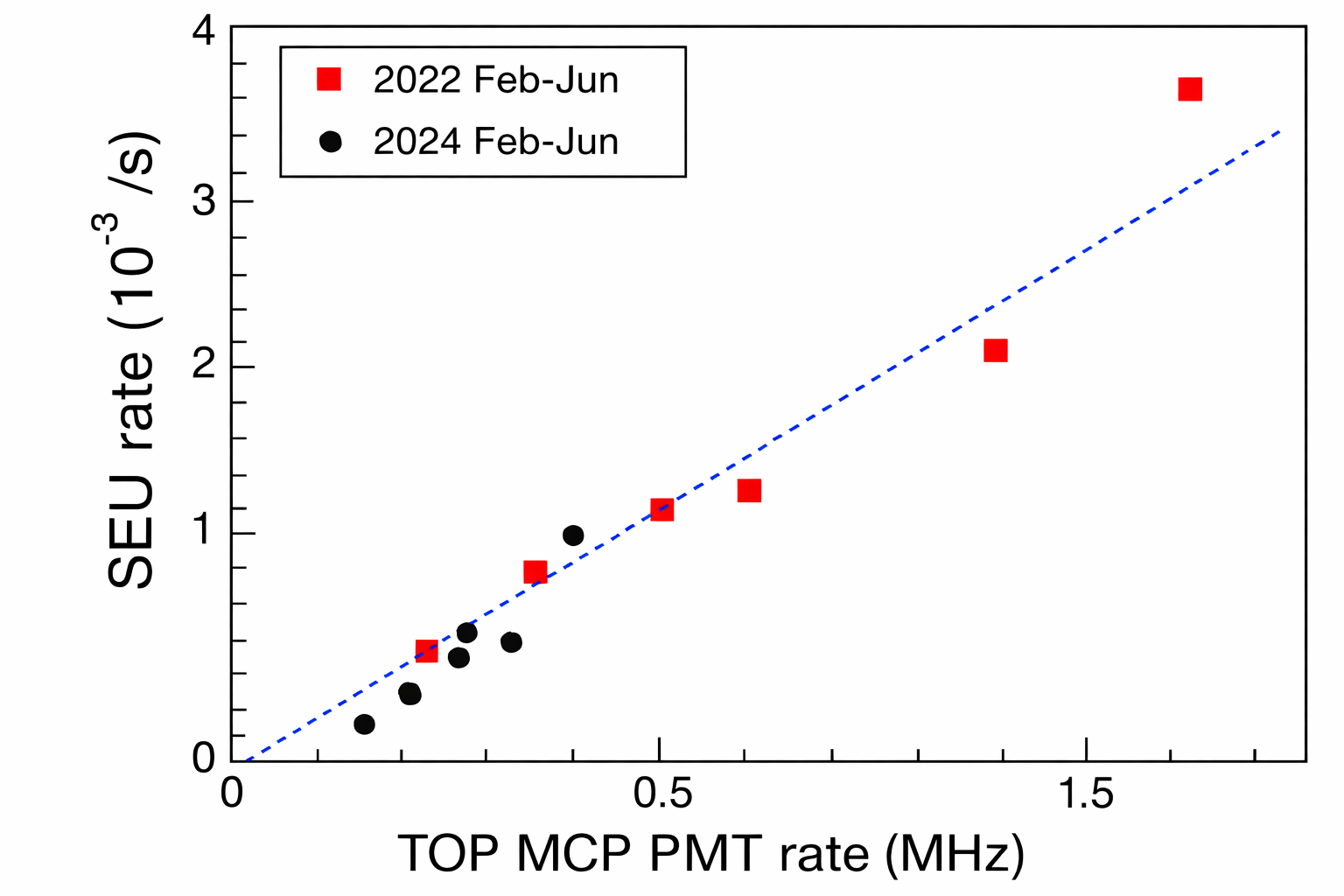}
    \caption{Relation between the SEU rate in ARICH and the hit rate of an MCP-PMT in the TOP detector. Each point corresponds to a period of SuperKEKB operation (typically 2 weeks) in 2022 and 2024. The dashed line is the fit result to a linear function. The TOP MCP-PMT rate is known to correlate well with the SuperKEKB beam background level.}
    \label{figure:seu}
\end{figure}

The data acquisition controller implements a process that controls the parameter settings of the readout cards and monitors their basic functionality, including readings of supply voltages, temperatures, and single-event upset (SEU) counts (Figs.~\ref{figure:temperature}  and \ref{figure:seu}). It also monitors the temperature of the inlet and outlet water cooling pipes, as well as the status of the cooling unit. Additional temperature sensors at various parts of the detector are readout from a common Belle II environmental monitoring system.

\section{Commissioning and Operation}

\subsection{Commissioning}

The ARICH detector commissioning phase began early in 2018, concurrent with the commissioning of all other  Belle II spectrometer components. During this phase, the detector operated reliably. 

\begin{figure}[tbp]
	\centering
	\includegraphics[width=0.48\columnwidth]{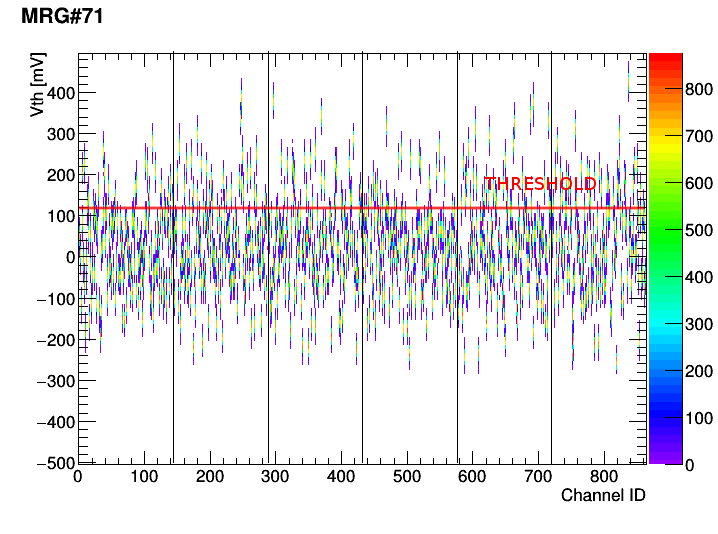}
	\includegraphics[width=0.48\columnwidth]{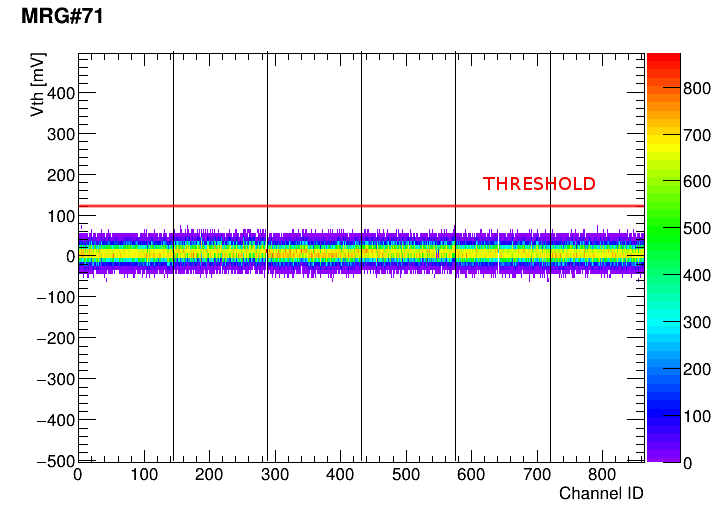}
	\caption{Response of the detector, a threshold scan for each of the channels of a single merger board with six front-end boards connected (corresponding to $6 \times 144 = 864$ channels). Left plot: non-calibrated offsets, right plot: threshold scan with calibrated offsets. 
    }
	\label{figure:thscan}
\end{figure}

Before the start of data acquisition, the front-end boards had to be programmed to set a common gain, discrimination threshold, and channel-dependent offsets for each ASIC. An example of a calibration of the channel offsets is shown in Fig.~\ref{figure:thscan}. Offsets calculated from the left plot were uploaded to the detector, allowing one common threshold to discriminate between hit and non-hit channels.
%

Initial commissioning revealed that the original cooling system required an upgrade to enhance its performance through more efficient thermal coupling to the read-out electronic boards, resulting in the system discussed in this paper. This upgrade was carried out during the summer shutdown in 2018. By the end of 2018, the commissioning phase was completed and was followed by the first physics data-taking runs in the spring of 2019.

\subsection{Operation}
\label{sec:operation}

The ARICH detector ran stably from the commissioning phase onward. As previously discussed, various parameters are continuously monitored to ensure stable performance and react in the event of performance deviations. 


To monitor the performance of the photo-sensors, light ($\lambda = 470$~nm) emitted from LEDs and distributed through optical fibers to 90 points in the gaps between the photo-sensors is used on a regular basis~\cite{HATAYA2017176}. The light is scattered from the aerogel tiles and is spread over the surface covered by HAPDs; it is used to identify noisy and dead channels, as well as to measure the relative sensitivity and gain of each channel.

\begin{figure}[tbp]
    \centering
    \includegraphics[width=0.3\columnwidth]{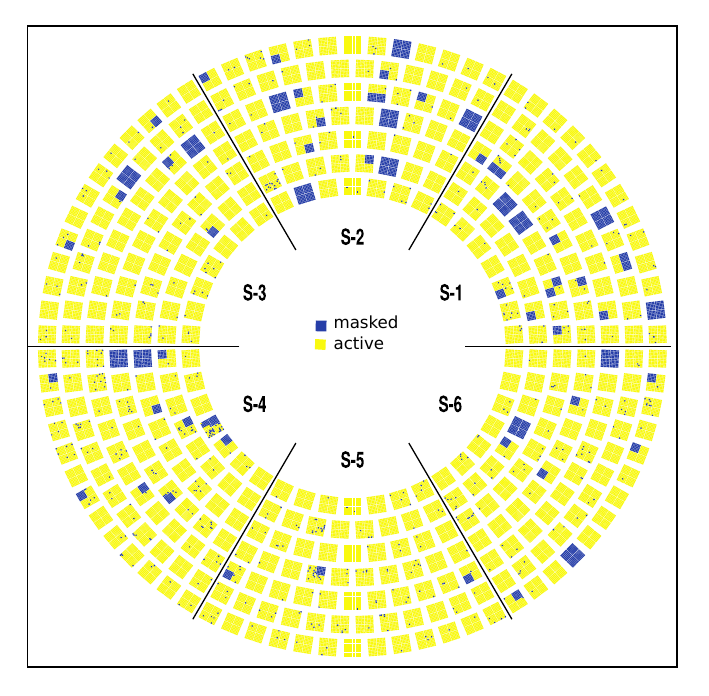}
    \includegraphics[width=0.60\columnwidth]{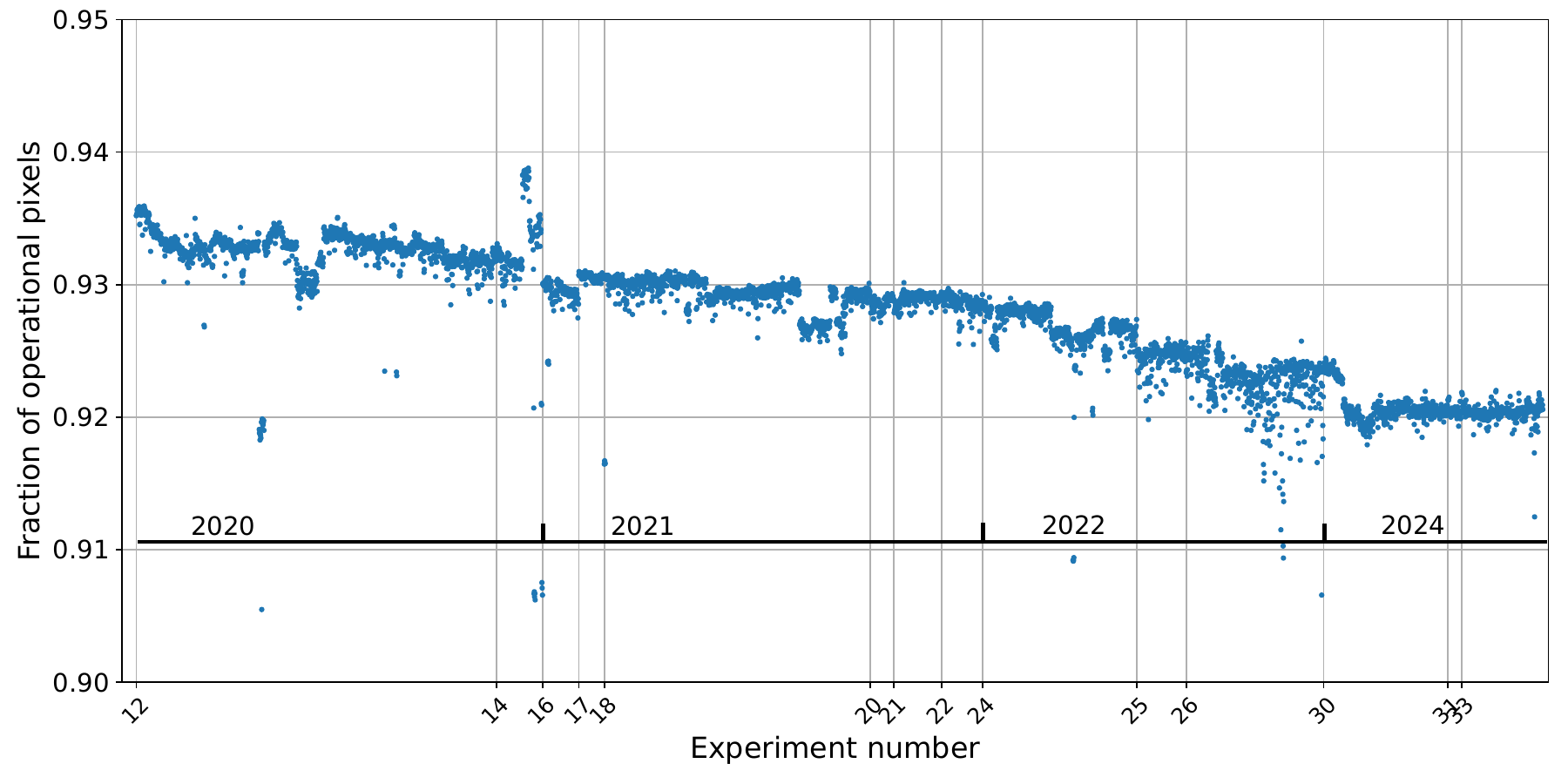}
    \caption{Masked channels: distribution over the photon detector at the beginning of Run 2 in December 2023 (left); variation of the number of operational channels with time (right).}
    \label{figure:masked}
    \end{figure}

\begin{figure}[tbp]
	\centering
	\includegraphics[width=0.64\columnwidth]{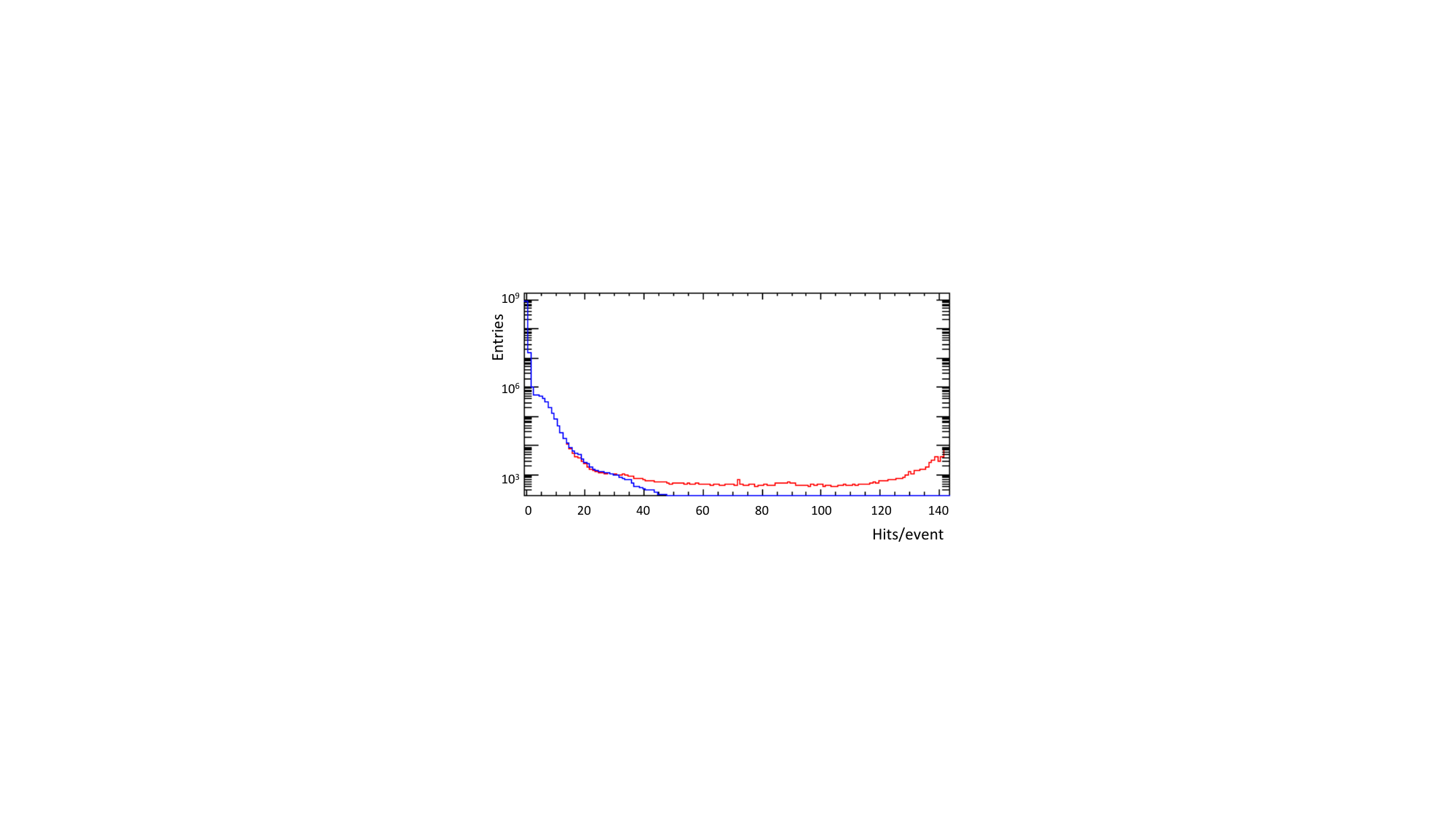}
	\caption{Number of hits per sensor per event. All sensor hits (red) and accepted hits (blue).}
	\label{figure:hotdead}
\end{figure}

During the acquisition of beam collision data, the quality of the data is constantly monitored. In the event-based filter, HAPD sensors with too many hits are rejected. Checks of dead and noisy channels (Figs.~\ref{figure:masked} and \ref{figure:hotdead}) are carried out for each run, and new channels that appear in this list are taken into account in the reconstruction. 

\begin{figure}[tbp]
    \centering
    \includegraphics[width=0.48\columnwidth]{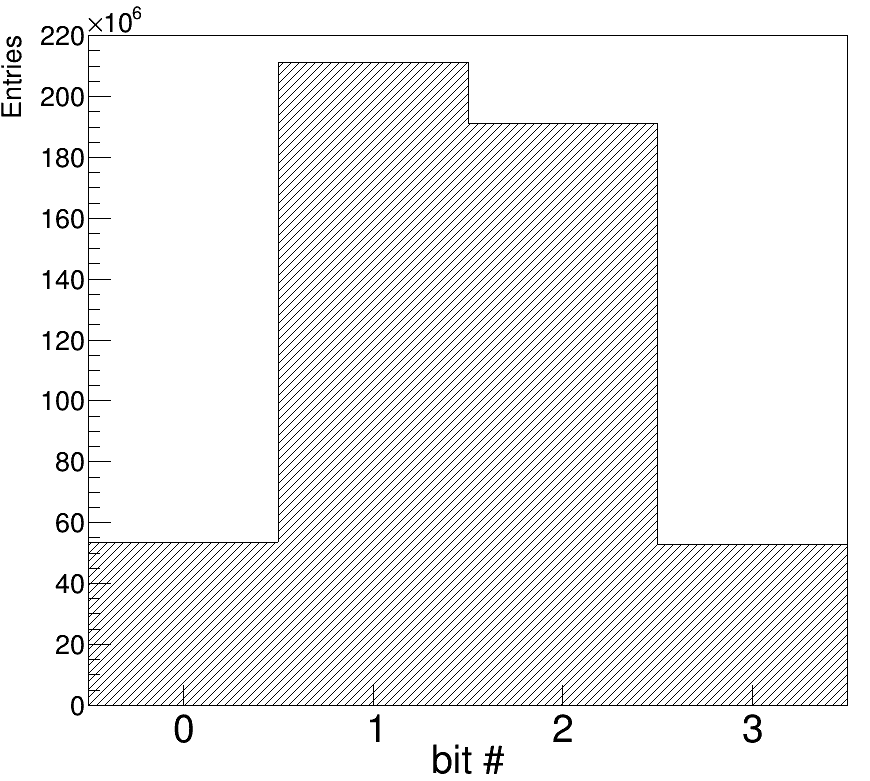}
    \caption{Timing alignment of hits: central two 125.6~ns long time intervals correspond to hits in the signal windows, while the left and right intervals correspond to off-time background hits.}
    \label{figure:time-alignment}
\end{figure}

As discussed in Sec.~\ref{sec:readout}, the hit information is read out in four adjacent $125.6~\mathrm{ns}$ long time intervals, where the middle two are adjusted to the correct timing with respect to the L1 trigger, and the two side bins are used to estimate the background level (Fig.~\ref{figure:time-alignment}). 

\begin{figure}[tbp]
    \centering
    \includegraphics[width=0.43\columnwidth]{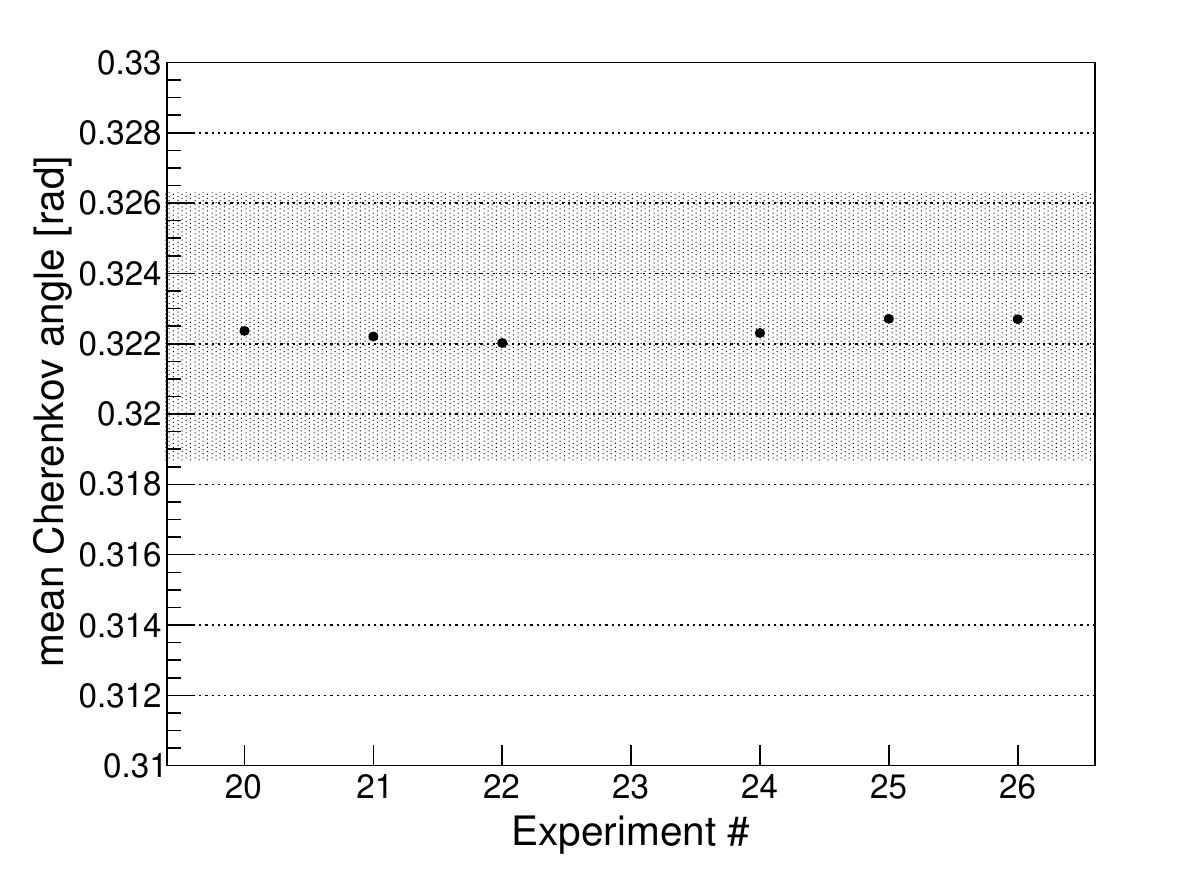}
    \includegraphics[width=0.48\columnwidth]{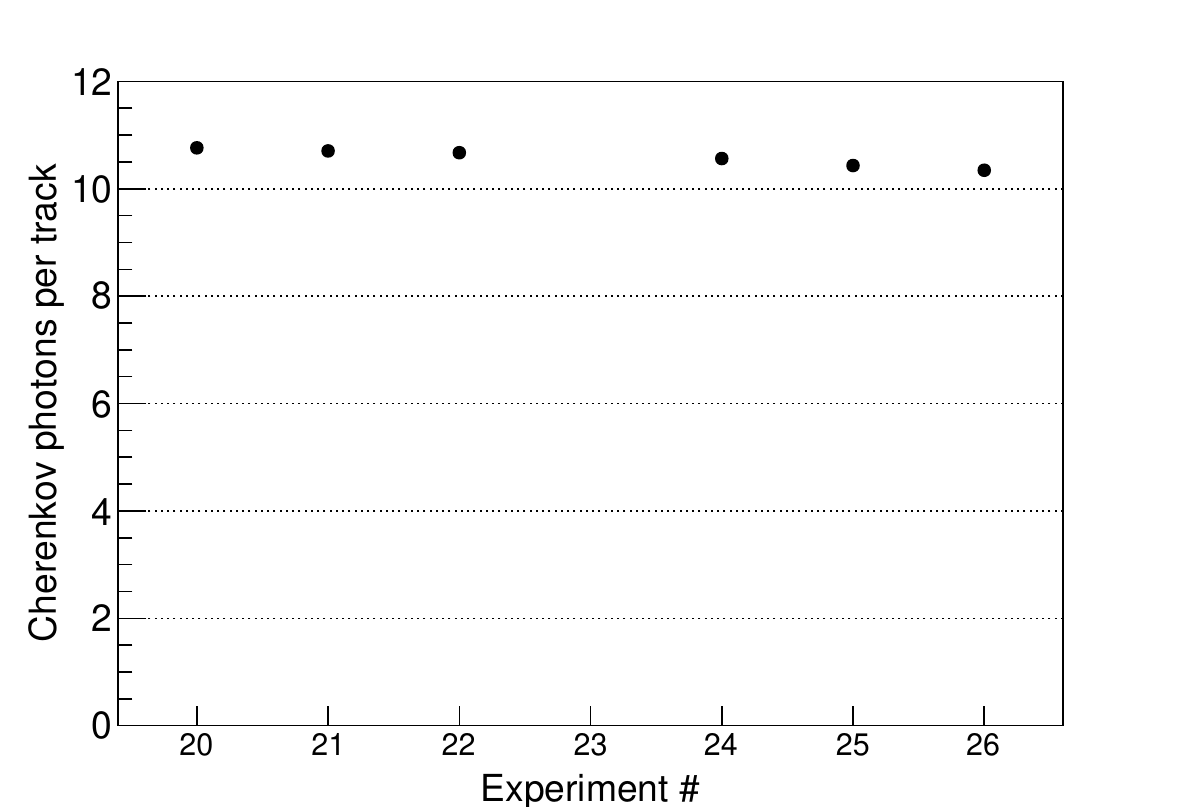}
    \caption{Detector performance as a function of time: Cherenkov angle (left) and number of photons (right) for ultra-relativistic muon tracks ($p > 4$~GeV/$c$); the gray region indicates the $\pm 1 \sigma_{\rm track}$ band for 4~GeV/$c$ pions.}
    \label{figure:perf-vs-time}
\end{figure}

We also monitor the overall performance of the detector by checking the two most relevant variables, the Cherenkov angle and the number of detected Cherenkov photons of ultra-relativistic muon tracks ($p > 4$~GeV/$c$). As shown in Fig.~\ref{figure:perf-vs-time}, the detector indeed performs reliably over extended periods of time.

\section{Simulation and reconstruction}

%

\begin{figure}[tbp]
    \centering
    \includegraphics[width=0.8\columnwidth]{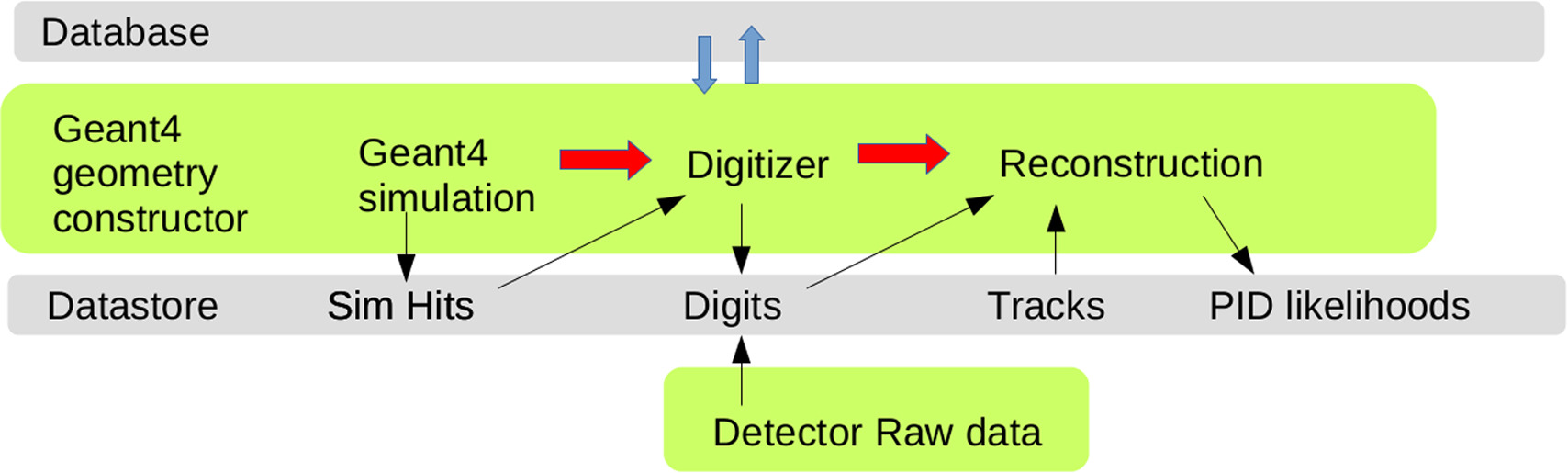}
    \caption{ARICH software scheme.}
    \label{fig:arich_software}
\end{figure}

The ARICH simulation and reconstruction software is integrated into Basf2 (Belle Analysis Software Framework 2)~\cite{Santelj:2017eum}, which provides a common framework for the Belle II detector simulation, event reconstruction, and data analysis. The code is organized in the form of independent modules (mostly written in C++) that perform a specific task and are included in the main event processing loop using Python-based steering files. For the ARICH detector, the basic structure of the event loop is shown in Figure \ref{fig:arich_software}. For the simulated data, the loop starts with the Geant4 simulation of the event. In the simulation, photon hits on the photon detectors are recorded. At the stage of digitization, these hits are converted into {\it digits}, which correspond to the actual output from the detector (i.e., containing only hit channel numbers). For the analysis of measured data, the event loop starts with the collection of digits, obtained directly from the raw data. Finally, the ARICH reconstruction module uses the collection of tracks impacting on the aerogel (obtained from the Belle II tracking system) and the collection of photon hits to calculate the PID likelihoods for each track.

\label{sec:software}

\subsection{Geant4 simulation}

The geometry of the ARICH detector is implemented in detail in the Geant4 framework. This includes all the main components relevant to the emission, propagation, and detection of Cherenkov photons, as well as the detector support structures, cooling system, and neutron shielding and cabling material. Since the detector geometry is described elsewhere, we only address here items particular to the simulation, which are also relevant when comparing the measured and simulated data.  

\paragraph{Aerogel plane}

In the simulation, all aerogel tiles of one layer have equal optical properties and thickness. The refractive indices of 1.045 and 1.055, and transmission lengths of 45 mm and 35 mm, are used for all the tiles in the upstream and downstream layers, respectively (the values given are at 400 nm, but their full wavelength dependency as measured in the QA tests is implemented in the simulation). The mounting structure of the tiles is assumed to be as in the real detector. The black paper that covers the tile slots is, for the sake of simplicity, not included in the simulation; its effect on photon absorption is mimicked by the unspecified optical properties of the aluminum, which results in optical photons being absorbed upon hitting it. 

\paragraph{Photon detector - HAPD}

In the simulation, we model the photon detector module as a ceramic box, featuring the quartz window on top, a vacuum inside, and a silicon APD sensor at the bottom. The dependency of the quartz refractive index on wavelength is implemented using data from the literature. The emission of photoelectrons from the photocathode and their propagation to the APDs is not included in the simulation. Instead, for the optical photons that hit the bottom surface of the quartz window, onto which the photocathode is coated, the quantum efficiency curve is applied. 
If the photon is registered as detected, it is removed, and its position is passed to the {\it digitizer} for the next steps of processing (as described in the following subsection). On the other hand, if the photon is not detected, it is left to propagate further, either by being internally reflected in the quartz window or by entering into the HAPD, where it can be either reflected from the APD back to the photocathode, or absorbed. The reflectivity of the APD surface in the simulation is wavelength-independent and is adjusted so that the fraction of reflected photons matches the one observed in the measured data. At this point, the same Q.E. curve is used for all HAPDs, which is corrected to the exact values during the digitisation step (see below).

\paragraph{Other components}
Here we provide a few relevant comments on other detector components. The mirrors are implemented as quartz planes coated with a reflective metal, where the reflectivity and its wavelength dependency are set to the values obtained in QA measurements. The front-end electronics (HAPD front-end boards and merger boards) are implemented as simple boxes, of correct dimensions, made of material commonly used for effective PCB description. The geometry of cooling bodies (behind each FEB and merger board) and cooling pipes is implemented in fine detail. On the other hand, the detector cabling (HV, LV, read out) is simulated by placing a simple homogeneous thin plane with the amount and type of material that effectively describes the cabling material, behind the photon-detector plane. Finally, the borated polyethylene neutron shield volumes are included with precisely described shapes and sizes.    

Since the peak Q.E. of HAPDs is $\sim 40$\%, only this fraction of emitted optical photons is actually propagated in the simulation, while the rest are removed immediately at the time of emission. This allows to speed up the detector simulation without affecting its output.

\subsection{Digitization}

In the digitizer module, the photon hits obtained from the Geant4-based  simulation are converted into a data format equivalent to that of the measured data (i.e., a collection of channel numbers for hit APD pads). There are four processes that are performed at this stage:
\begin{itemize}
    \item Pixelization: based on the registered photon hit position, we calculate the pad number to which this position corresponds. Here, we assume that the emitted photoelectron travels only in the 
    $z$-direction, i.e., along the magnetic field lines. 
    \item Channel-by-channel Q.E. correction: at the simulation level, a common wavelength dependence of Q.E. is used for all HAPDs. However, since relatively large differences in the Q.E. are observed between individual samples, and the Q.E. non-uniformity over the surface of individual sensors cannot be neglected, we apply a channel-by-channel Q.E. correction, using the measured Q.E. surface maps of individual HAPDs.
    \item Dead channel masking: about 5\% of channels are non-operational; we maintain a list of these channels in the database and utilize it for the production of simulated data.
    \item Effect of negative polarity cross-talk: if a single pad is hit by a large number of photoelectrons, the efficiency of neighbouring pads to detect photoelectrons is reduced. This effect is included in the digitizer, where we count the number of photoelectrons on each pad and proportionally lower the efficiency of the neighbouring pads. 
\end{itemize}

The list of digits, which contains the photon-detector module ID number, channel ID number, and a hit bitmap\footnote{These are four time-consecutive bits, which indicate whether the signal exceeded the threshold value in a given time bin or not.} is the final output of the Geant4 simulation.   

The simulation of ARICH detector response is checked using high momentum muon tracks from $e^+e^- \to \mu^+\mu^-$ data, comparing the Cherenkov ring image in measured and simulated events.  
\begin{figure}[tbp]
\begin{center}
\includegraphics[width=0.8\columnwidth]{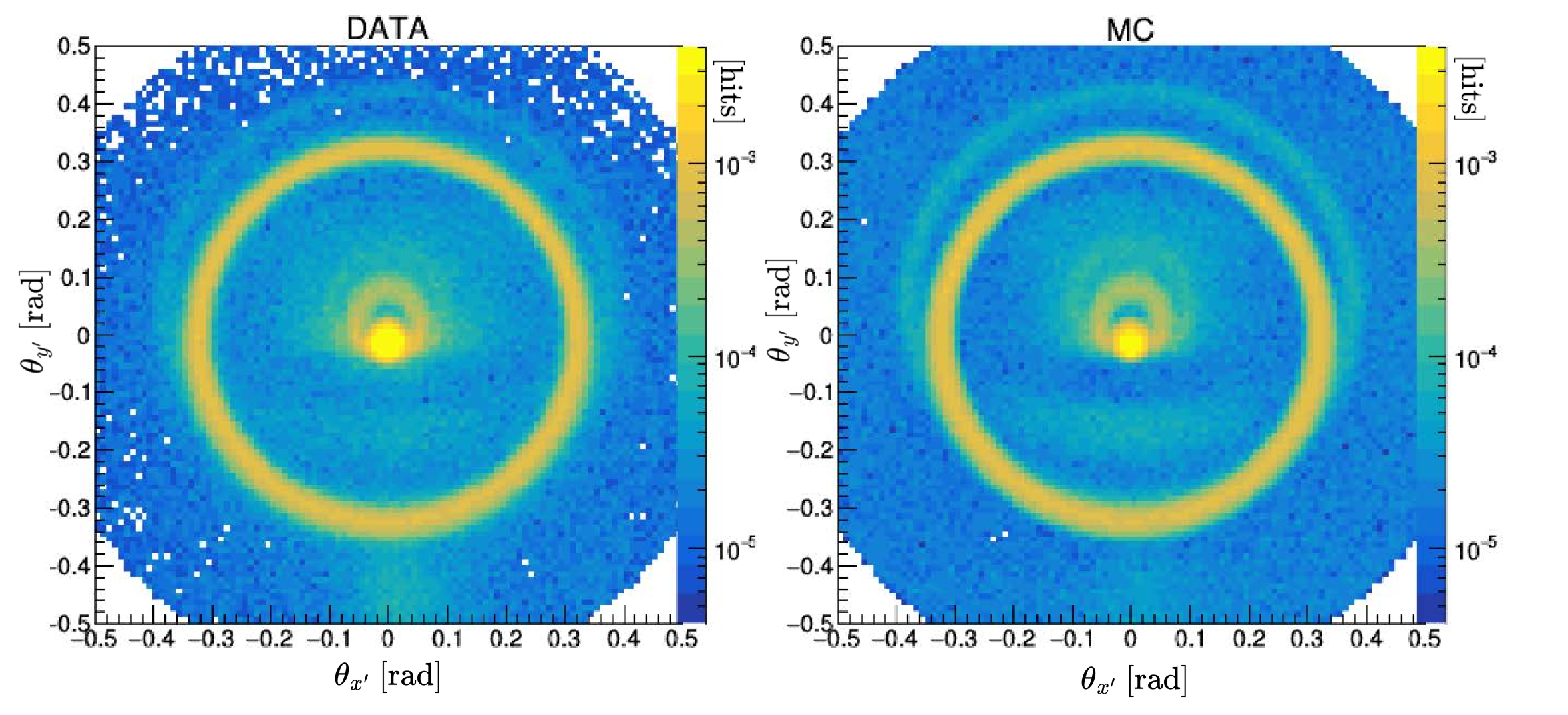}
\includegraphics[width=0.8\columnwidth]{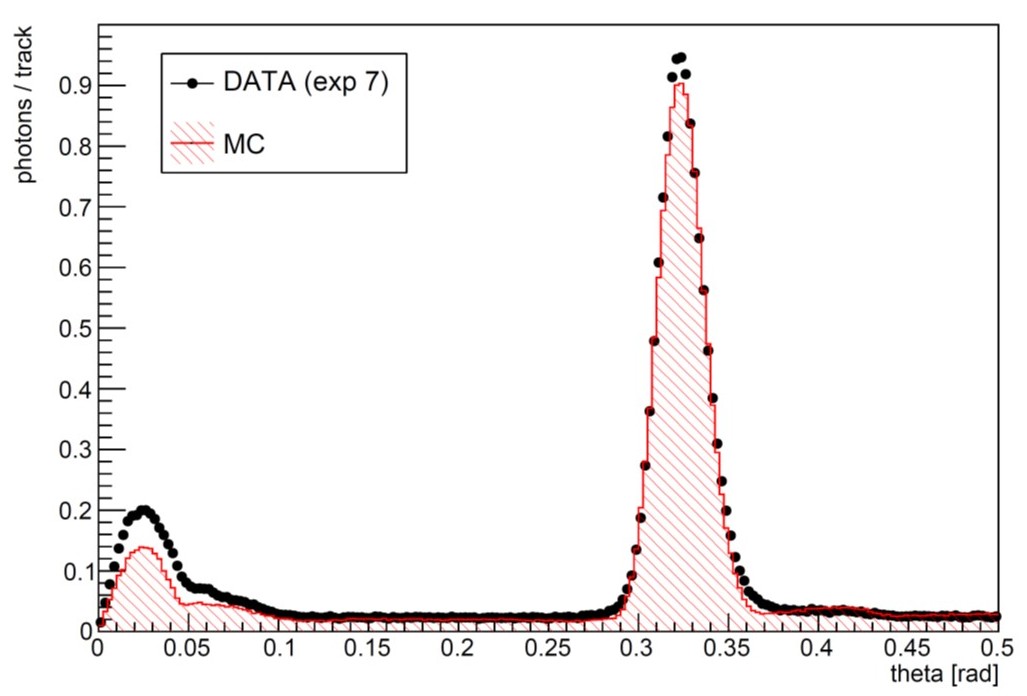}
\end{center}
\caption{Calibration of the simulated detector response; top: comparison of accumulated Cherenkov ring images as observed in the measured data and the detector simulation; bottom: comparison of measured and simulated Cherenkov angle distributions (integral of top two plots over the ring azimuthal angle).
}
\label{fig:richrings}
\end{figure}
The top two plots of Fig.~\ref{fig:richrings} show the reconstructed Cherenkov ring image in the angular coordinate 
system of a track\footnote{In this system, the track direction corresponds to ($\theta_{x'}=0$,\ $\theta_{y'}=0$), and the 
distance from the origin shows the photon direction polar angle, i.e., Cherenkov angle.} for 
the measured and simulated data. The image is obtained as a normalized sum of Cherenkov rings 
from $\mathcal{O}$(10k) muon tracks. Several detailed features, such as Cherenkov photons 
produced in the quartz window of photon detectors and an "echo ring" originating from the non-converted photons reflected from the APD surface back to the photocathode, are 
reproduced well in the simulation~\cite{Santelj:2023iim}. Some discrepancies between the measured and simulated 
data are visible in the peak produced by charged tracks in the window of photon detectors at small Cherenkov angles. This discrepancy is due to the difficulty of correctly modeling the optical properties of the photocathode and the response of the avalanche photodiode to a large number of photons. Based on 
these observations, the detector simulation will be adjusted.

\subsection{Event reconstruction} 

Reconstructed tracks from the Belle II tracking system are extrapolated to the ARICH detector volume. For those that pass through the aerogel layer, a likelihood function is constructed for each of the six different particle type hypotheses ($e,\mu,\pi,K,p,d$). The likelihood function compares the observed pattern of photon hits with that expected for the given particle type hypothesis and track parameters (position, direction, and momentum on the aerogel plane) as obtained from the track extrapolation. 

The likelihood function is constructed as the product of the probabilities for individual pads to record the observed number of hits — either zero or one, since single and multiple photon hits are not distinguished — under a given particle-type hypothesis. The probability of pad $i$ being hit by $m_i$ photoelectrons is given by the Poissonian distribution, i.e., $p_i=e^{-n_i}n_i^{m_i}/m_i!$, where $n_i$ is the number of photoelectrons expected to hit the pad $i$. The value of $n_i$ is the sum of the expected number of hits from two radiator layers and from the background. 
The contribution of each radiator to the signal on the pad  $i$ is calculated according to the probability distribution in the polar and azimuthal angles relative to the track direction over the solid angle subtended by the pad. The method is discussed in detail in \cite{Pestotnik:2020calib}.

Following P.~Baillon~\cite{baillon} and R.~Forty~\cite{forty} for the construction of the likelihood function, we first note that the probability of a pad being fired or not is $1-e^{-n_i}$ and $e^{-n_i}$, respectively, and the logarithm of the likelihood function is

\begin{align*}
\ln \mathcal{L} = \sum_{hit}\ln{(1-e^{-n_i})} + \sum_{no~hit}\ln{(e^{-n_i})} \\ = \sum_{hit}\left[ n_i + \ln{(1-e^{-n_i})}\right] - \sum_{all} n_i  =\sum_{hit}\ln{(e^{n_i}-1)} - N,
\end{align*}

\noindent where $hit~/~ no~hit~/~all$ indicates the subset of pads included in the sum, and we note that the sum of $n_i$ over all pads is equal to the total number of photons expected to be detected (denoted by $N$). Instead of having to obtain $n_i$ for all pads, evaluating the likelihood in this form reduces to calculating $n_i$ only for pads that registered a hit, and to estimating $N$. Both $n_i$ and $N$ have to be evaluated for each particle type hypothesis. Finally, the difference of likelihoods of two hypotheses (e.g. $\ln \mathcal{L}^{\pi} - \ln \mathcal{L}^{K}$) is used for separation between particle species ($\pi$, $K$).

 \subsection{Calibration and alignment} 
 
To optimize the Cherenkov angle resolution, the ARICH detector must be aligned with the Belle II tracking system; additionally, the orientation of the planar mirrors must be determined. This is achieved by using recorded data, primarily from \(e^+e^- \to \mu^+\mu^-\) events~\cite{Pestotnik:2020calib, Tamechika:2020zzg}. We maximize the muon identification likelihood by varying the positions and orientations of ARICH components. 

We then determine the parameters of the probability density function used for evaluation of the expected number of photons per pad (\( n_i \)) in the likelihood function, which incorporates both background and signal contributions. The signal contribution is determined by the Cherenkov angle resolution (\( \sigma_{\theta} \)) and the total number of photons emitted in the aerogel by a $\beta=1$ particle. The Cherenkov angle distribution for saturated rings of muons from \(e^+e^- \to \mu^+\mu^-\) events is used to determine these parameters, along with the expected background hit level. Besides these calibrations, which are very stable and nominally performed only at the beginning of each data-taking period, we produce a list of dysfunctional (dead or noisy) channels for each data-taking run (typically a few hours long). The list is based on the analysis of channel occupancy data, and the procedure to create it is fully automated. This information is used to properly account for non-active regions of the detector in the likelihood function.

Parameters from all stages of calibration are stored in a central database of the experiment and accessed by the reconstruction algorithm from there.

\section{Particle identification performance}

The ARICH detector ran stably throughout all running periods of the Belle II experiment. The fraction of dead or noisy channels has been at the 5\% level; as determined from simulation, such a level of missing channels is far below the level that would impact the particle identification performance of the device. 

The performance of the detector was first checked using a large number of available muon tracks originating from the $e^+e^- \to \mu^+\mu^-$ data. Muons from these events have the energy of about 7~GeV, producing essentially saturated Cherenkov rings in the ARICH. From the bottom plot of Fig.~\ref{fig:richrings}, we determine the number of signal Cherenkov photons 
and the Cherenkov angle resolution for measured data and simulation by fitting both
distributions with a single Gaussian function for the peak and a first-order polynomial for 
the background distribution. The number of signal photons per muon track is $N_{sig}
^{data} = 11.38$ and $N_{sig}^{MC} = 11.27$ for the measured and simulated data respectively, 
while the corresponding Cherenkov angle resolutions (i.e., signal peak width) are $
\sigma_c^{data} = 12.70 \textrm{ mrad}$ and $\sigma_c^{MC} = 12.75\textrm{ mrad}$ (with 
negligible uncertainties from the fit). 

\begin{figure*}[tbp]
\begin{center}
\includegraphics[width=0.85\textwidth]{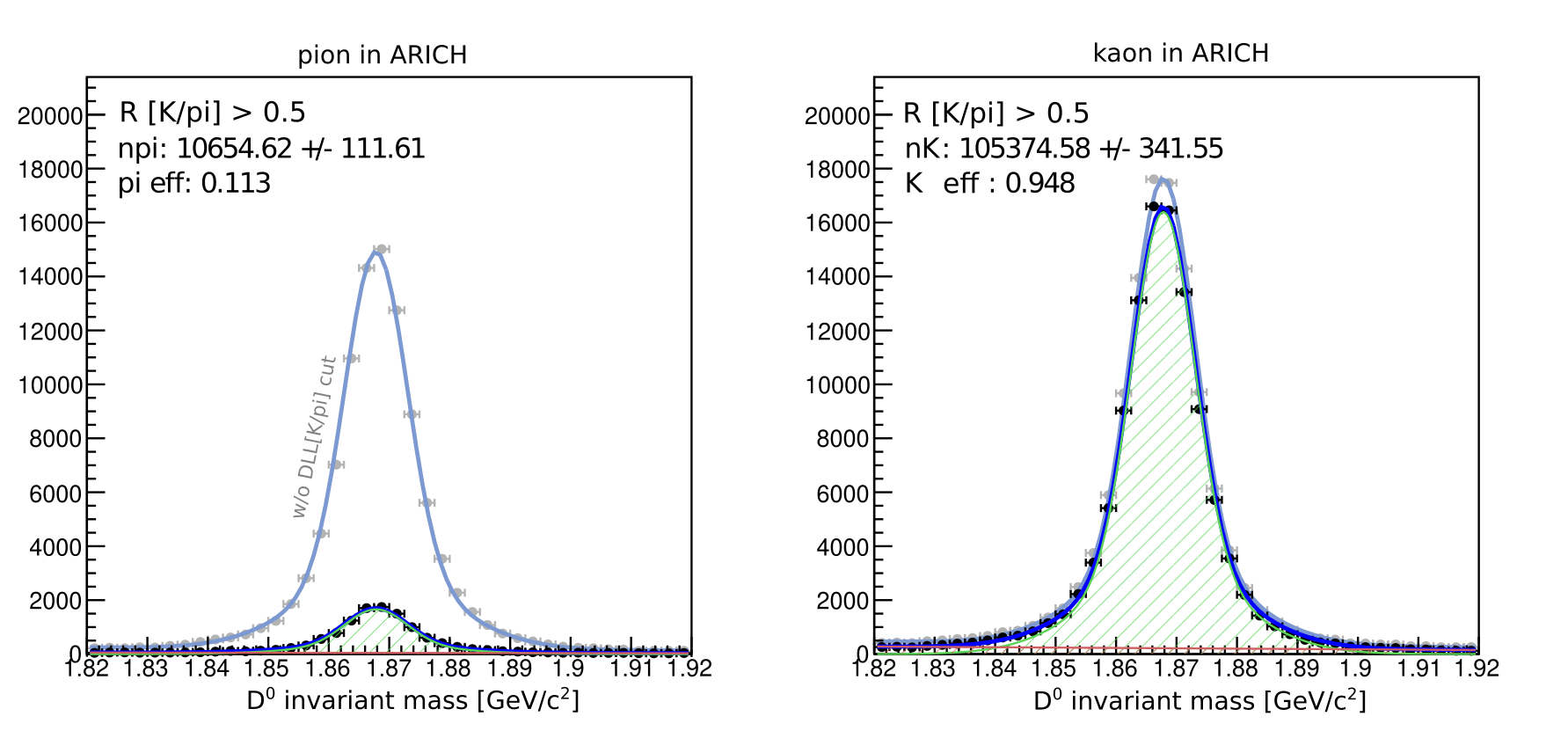}
\includegraphics[width=0.48\textwidth]{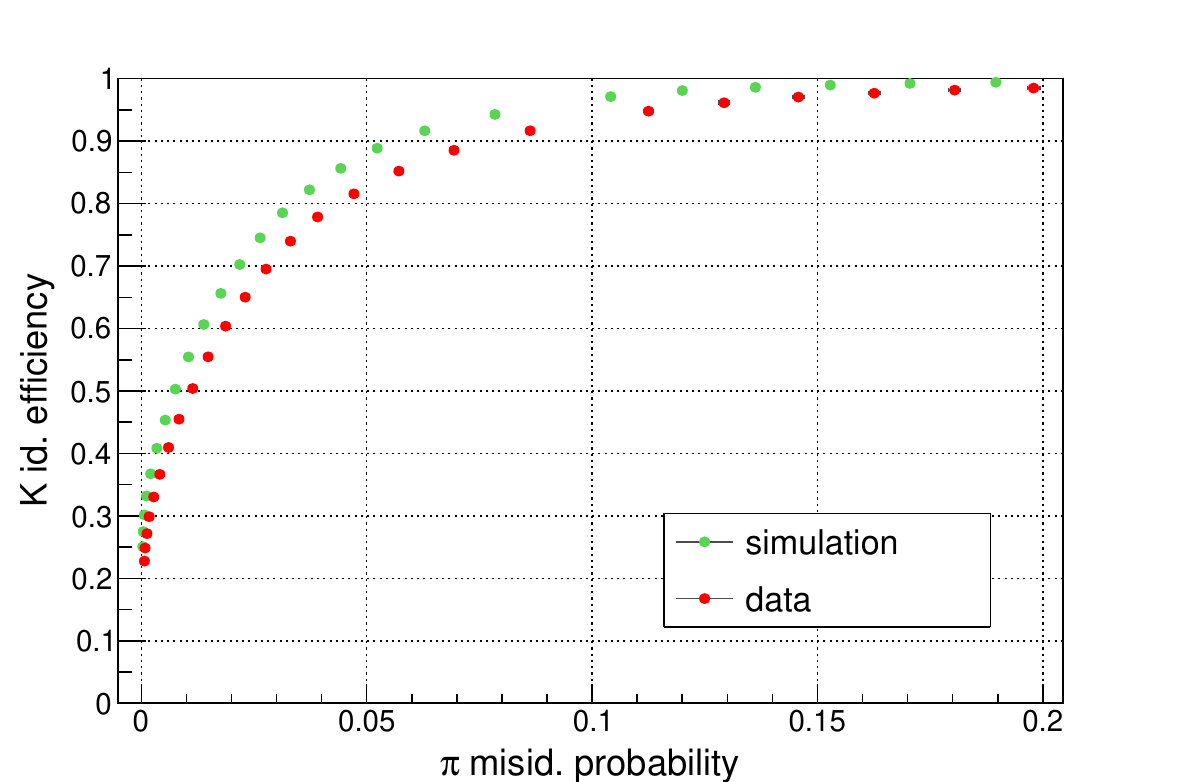}
\includegraphics[width=0.48\textwidth]{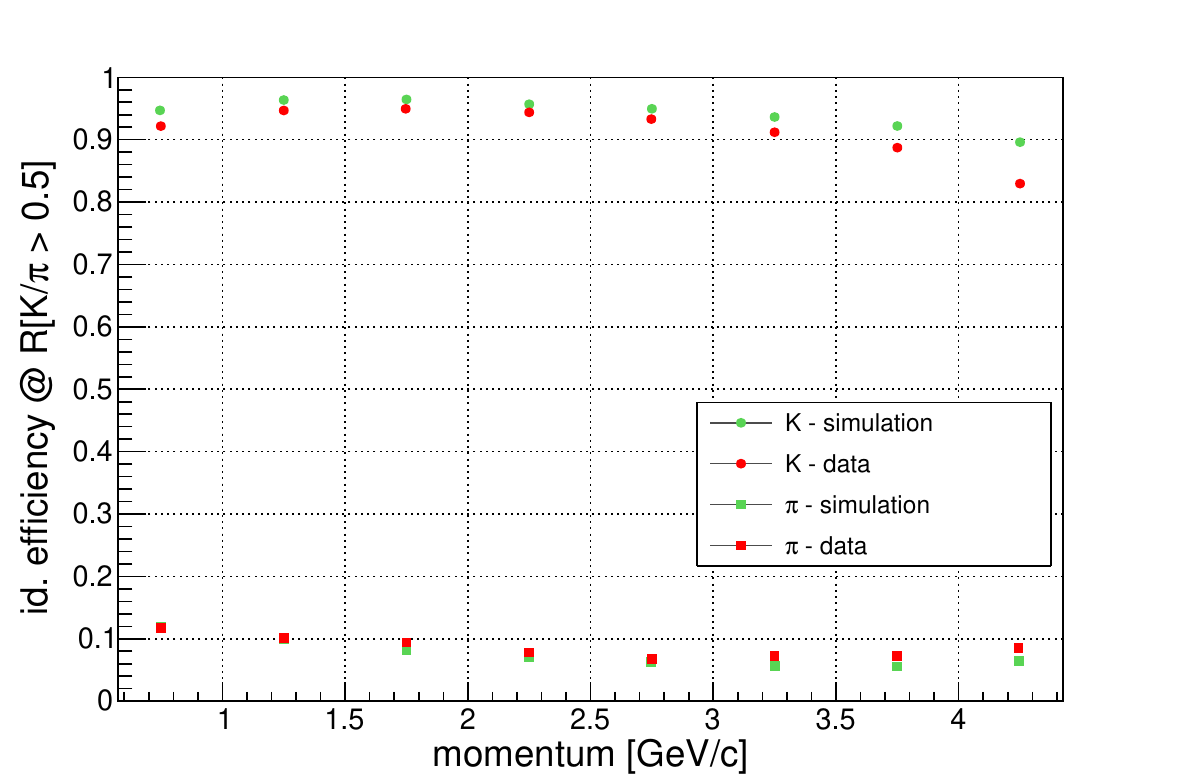}
\end{center}
\caption{ARICH detector, performance. Top: examples of fits of the 
$K^{\mp}\pi^{\pm}$ invariant 
mass distributions from which we determine the kaon identification efficiency and pion 
misidentification probability at a given criterion on $R[K/\pi]$. Bottom left:  kaon identification 
efficiency versus pion misidentification probability for all kaon/pion tracks from $D^0$ decays that 
enter the ARICH detector; right: kaon identification efficiency and pion misidentification probability as a function of track momentum for a fixed likelihood ratio, $R[K/\pi]>0.5$.
}
\label{fig:richperf}
\end{figure*}

We estimate the ability of the  detector to discriminate between pions and kaons using pion and kaon 
tracks from $D^{*+}\to D^0[\to K^-\pi^+]~\pi^+_{\textrm{slow}}$ and $D^{*-}\to \bar{D}^0[\to
K^+\pi^-]~\pi^-_{\textrm{slow}}$ decay chains in data. 
Here, pion and kaon tracks can be identified 
independently of ARICH information, based on their charge in association with the charge of
low-momentum pion ($\pi^{\pm}_{\textrm{slow}}$) from $D^{*\pm}$. In addition, the background
level in this decay mode can be effectively reduced by requiring the difference of 
reconstructed invariant masses of $D^*$ and $D^0$ mesons to be within a narrow window around 
the expected value ($|M_{D^*} - M_{D^0}-145.43 \textrm{ MeV}/c^2| < 1.5 \textrm{ MeV}/c^2$ is 
used). We determine the efficiency of kaon identification
and pion misidentification probability
from the signal yield of reconstructed $D^0$ mesons, where pion and kaon 
tracks used in the reconstruction are required to satisfy the imposed selection criteria on the PID 
likelihood ratio $R[K/\pi]$ obtained from the ARICH detector, with $R[K/\pi]=\mathcal{L}_K/(\mathcal{L}_K+\mathcal{L}
_{\pi})$. The $D^0$ signal yield is determined from the fit of the $D^0$ invariant mass 
distribution using a Gaussian function for the signal and a constant value for the background 
distributions. An example of such a fit is shown in the top two plots of figure \ref{fig:richperf}.

The left  (right) plot shows data distribution and fit of $D^0$ invariant mass for candidates with a pion 
(kaon) track entering ARICH. The light grey points and curves show the case when no selection is 
imposed on $R[K/\pi]$, and solid points and curves after imposing a $R[K/\pi]$ requirement. 
Repeating 
the fit using different $R[K/\pi]$ criteria results in the bottom left plot of Fig.~\ref{fig:richperf}, which shows the kaon identification efficiency at different pion 
misidentification probabilities. The right bottom plot shows the dependence of these two quantities on the track momentum, where a fixed criterion of $R[K/\pi]>0.5$ is used at all points. The performance is slightly worse than expected from the simulations (up to a few \% in K identification efficiency). The main contributions to this discrepancy originate from tracks that enter aerogel close to the tile boundary and from particles that get scattered before reaching the aerogel radiator~\cite{Santelj:2023iim}; further improvements to take this into account are under study. 
In a similar manner, $\Lambda \to p \pi^-$ decays are used to verify the pion-proton separation; at 4~GeV/$c$, the proton identification efficiency is 91\% with a pion misidentification probability of 8\%.  
We note that the combined hadron identification performance of all subsystems in the Belle II spectrometer is discussed in detail in a dedicated  report~\cite{Belle-II:2025tpe}. To summarize, the ARICH detector provides excellent pion-kaon, pion-proton, and kaon-proton discrimination, 
and at momenta below 1~GeV/$c$, a modest discrimination between pions, muons, and electrons.


\section{Summary}

The Aerogel RICH detector in the forward end-cap of the Belle II spectrometer is an integral part of the Belle II particle identification system. It is based on a novel multi-layer radiator configuration and utilizes a new type of single-photon sensor, hybrid avalanche photodetectors (HAPD). It provides excellent discrimination between pions and kaons in the full kinematic range of the experiment (up to $4$ GeV/$c$). The detector performed very well throughout the Belle II data-taking and is expected to continue contributing to the future physics harvest of the experiment.

\section{Acknowledgements}

We thank the SuperKEKB group for the excellent operation of the accelerator, the KEK cryogenics group for the efficient operation of the solenoid, the KEK computer group for on-site computing support, and the raw-data centers at BNL, DESY, GridKa, IN2P3, and INFN for off-site computing support. 

This work was supported by the following funding sources: the Higher Education and Science Committee of the Republic of Armenia Grant No. 23LCG-1C011; the European Research Council, Horizon 2020 ERC-Advanced Grant No. 884719; the Japan Society for the Promotion of Science KAKENHI Grants Number JP24244035, JP18H05226, and the Ministry of Education, Culture, Sports, Science, and Technology (MEXT) of Japan; the Slovenian Research and Innovation Agency research grants No. J1-9124, J1-4358, J1-50010 and P1-0135.

\bibliographystyle{unsrt}
\bibliography{arich}

\end{document}